# A Measurement of the Proton Structure Function $F_2(x, Q^2)$

H1 Collaboration

**Abstract:**

A measurement of the proton structure function $F_2(x, Q^2)$ is reported for momentum transfers squared $Q^2$ between 4.5 GeV$^2$ and 1600 GeV$^2$ and for Bjorken $x$ between $1.8 \cdot 10^{-4}$ and 0.13 using data collected by the HERA experiment H1 in 1993. It is observed that $F_2$ increases significantly with decreasing $x$, confirming our previous measurement made with one tenth of the data available in this analysis. The $Q^2$ dependence is approximately logarithmic over the full kinematic range covered. The subsample of deep inelastic events with a large pseudo-rapidity gap in the hadronic energy flow close to the proton remnant is used to measure the "diffractive" contribution to $F_2$.

# H1 Collaboration

T. Ahmed[3], S. Aid[13], A. Akhundov[35,40], V. Andreev[24], B. Andrieu[28], R.-D. Appuhn[11], M. Arpagaus[36], A. Babaev[26], J. Baehr[35], J. Bán[17], P. Baranov[24], E. Barrelet[29], W. Bartel[11], M. Barth[4], U. Bassler[29], H.P. Beck[37], H.-J. Behrend[11], A. Belousov[24], Ch. Berger[1], H. Bergstein[1], G. Bernardi[29], R. Bernet[36], G. Bertrand-Coremans[4], M. Besançon[9], R. Beyer[11], P. Biddulph[22], J.C. Bizot[27], V. Blobel[13], K. Borras[8], F. Botterweck[4], V. Boudry[28], A. Braemer[14], F. Brasse[11], W. Braunschweig[1], V. Brisson[27], D. Bruncko[17], C. Brune[15], R.Buchholz[11], L. Büngener[13], J. Bürger[11], F.W. Büsser[13], A. Buniatian[11,39], S. Burke[18], G. Buschhorn[26], A.J. Campbell[11], T. Carli[26], F. Charles[11], D. Clarke[5], A.B. Clegg[18], B. Clerbaux[4], M. Colombo[8], J.G. Contreras[8], C. Cormack[19], J.A. Coughlan[5], A. Courau[27], Ch. Coutures[9], G. Cozzika[9], L. Criegee[11], D.G. Cussans[5], J. Cvach[30], S. Dagoret[29], J.B. Dainton[19], M. Danilov[23], W.D. Dau[16], K. Daum[34], M. David[9], E. Deffur[11], B. Delcourt[27], L. Del Buono[29], A. De Roeck[11], E.A. De Wolf[4], P. Di Nezza[32], C. Dollfus[37], J.D. Dowell[3], H.B. Dreis[2], V. Droutskoi[23], J. Duboc[29], D. Düllmann[13], O. Dünger[13], H. Duhm[12], J. Ebert[34], T.R. Ebert[19], G. Eckerlin[11], V. Efremenko[23], S. Egli[37], H. Ehrlichmann[35], S. Eichenberger[37], R. Eichler[36], F. Eisele[14], E. Eisenhandler[20], R.J. Ellison[22], E. Elsen[11], M. Erdmann[14], W. Erdmann[36], E. Evrard[4], L. Favart[4], A. Fedotov[23], D. Feeken[13], R. Felst[11], J. Feltesse[9], J. Ferencei[15], F. Ferrarotto[32], K. Flamm[11], M. Fleischer[11], M. Flieser[26], G. Flügge[2], A. Fomenko[24], B. Fominykh[23], M. Forbush[7], J. Formánek[31], J.M. Foster[22], G. Franke[11], E. Fretwurst[12], E. Gabathuler[19], K. Gabathuler[33], K. Gamerdinger[26], J. Garvey[3], J. Gayler[11], M. Gebauer[8], A. Gellrich[11], H. Genzel[1], R. Gerhards[11], U. Goerlach[11], L. Goerlich[6], N. Gogitidze[24], M. Goldberg[29], D. Goldner[8], B. Gonzalez-Pineiro[29], I. Gorelov[23], P. Goritchev[23], C. Grab[36], H. Grässler[2], R. Grässler[2], T. Greenshaw[19], G. Grindhammer[26], A. Gruber[26], C. Gruber[16], J. Haack[35], D. Haidt[11], L. Hajduk[6], O. Hamon[29], M. Hampel[1], E.M. Hanlon[18], M. Hapke[11], W.J. Haynes[5], J. Heatherington[20], G. Heinzelmann[13], R.C.W. Henderson[18], H. Henschel[35], R. Herma[1], I. Herynek[30], M.F. Hess[26], W. Hildesheim[11], P. Hill[5], K.H. Hiller[35], C.D. Hilton[22], J. Hladký[30], K.C. Hoeger[22], M. Höppner[8], R. Horisberger[33], V.L. Hudgson[3], Ph. Huet[4], M. Hütte[8], H. Hufnagel[14], M. Ibbotson[22], H. Itterbeck[1], M.-A. Jabiol[9], A. Jacholkowska[27], C. Jacobsson[21], M. Jaffre[27], J. Janoth[15], T. Jansen[11], L. Jönsson[21], K. Johannsen[13], D.P. Johnson[4], L. Johnson[18], H. Jung[29], P.I.P. Kalmus[20], D. Kant[20], R. Kaschowitz[2], P. Kasselmann[12], U. Kathage[16], J. Katzy[14], H.H. Kaufmann[35], S. Kazarian[11], I.R. Kenyon[3], S. Kermiche[25], C. Keuker[1], C. Kiesling[26], M. Klein[35], C. Kleinwort[13], G. Knies[11], W. Ko[7], T. Köhler[1], J. Köhne[26], H. Kolanoski[8], F. Kole[7], S.D. Kolya[22], V. Korbel[11], M. Korn[8], P. Kostka[35], S.K. Kotelnikov[24], T. Krämerkämper[8], M.W. Krasny[6,29], H. Krehbiel[11], D. Krücker[2], U. Krüger[11], U. Krüner-Marquis[11], J.P. Kubenka[26], H. Küster[2], M. Kuhlen[26], T. Kurča[17], J. Kurzhöfer[8], B. Kuznik[34], D. Lacour[29], F. Lamarche[28], R. Lander[7], M.P.J. Landon[20], W. Lange[35], P. Lanius[26], J.-F. Laporte[9], A. Lebedev[24], C. Leverenz[11], S. Levonian[11,24], Ch. Ley[2], A. Lindner[8], G. Lindström[12], F. Linsel[11], J. Lipinski[13], B. List[11], P. Loch[27], H. Lohmander[21], G.C. Lopez[20], V. Lubimov[23], D. Lüke[8,11], N. Magnussen[34], E. Malinovski[24], S. Mani[7], R. Maraček[17], P. Marage[4], J. Marks[25], R. Marshall[22], J. Martens[34], R. Martin[11], H.-U. Martyn[1], J. Martyniak[6], S. Masson[2], T. Mavroidis[20], S.J. Maxfield[19], S.J. McMahon[19], A. Mehta[22], K. Meier[15], D. Mercer[22], T. Merz[11], C.A. Meyer[37], H. Meyer[34], J. Meyer[11], S. Mikocki[6], D. Milstead[19], F. Moreau[28], J.V. Morris[5], E. Mroczko[6], G. Müller[11], K. Müller[37], P. Murín[17], V. Nagovizin[23], R. Nahnhauer[35], B. Naroska[13], Th. Naumann[35], P.R. Newman[3], D. Newton[18], D. Neyret[29], H.K. Nguyen[29], T.C. Nicholls[3],

F. Niebergall[13], C. Niebuhr[11], R. Nisius[1], G. Nowak[6], G.W. Noyes[5], M. Nyberg-Werther[21], M. Oakden[19], H. Oberlack[26], U. Obrock[8], J.E. Olsson[11], E. Panaro[12], A. Panitch[4], C. Pascaud[27], G.D. Patel[19], E. Peppel[35], E. Perez[9], J.P. Phillips[22], Ch. Pichler[12], D. Pitzl[36], G. Pope[7], S. Prell[11], R. Prosi[11], G. Rädel[11], F. Raupach[1], P. Reimer[30], S. Reinshagen[11], P. Ribarics[26], H.Rick[8], V. Riech[12], J. Riedlberger[36], S. Riess[13], M. Rietz[2], E. Rizvi[20], S.M. Robertson[3], P. Robmann[37], H.E. Roloff[35], R. Roosen[4], K. Rosenbauer[1] A. Rostovtsev[23], F. Rouse[7], C. Royon[9], K. Rüter[26], S. Rusakov[24], K. Rybicki[6], R. Rylko[20], N. Sahlmann[2], E. Sanchez[26], D.P.C. Sankey[5], M. Savitsky[23], P. Schacht[26], S. Schiek[11], P. Schleper[14], W. von Schlippe[20], C. Schmidt[11], D. Schmidt[34], G. Schmidt[13], A. Schöning[11], V. Schröder[11], E. Schuhmann[26], B. Schwab[14], A. Schwind[35], U. Seehausen[13], F. Sefkow[11], M. Seidel[12], R. Sell[11], A. Semenov[23], V. Shekelyan[23], I. Sheviakov[24], H. Shooshtari[26], L.N. Shtarkov[24], G. Siegmon[16], U. Siewert[16], Y. Sirois[28], I.O. Skillicorn[10], P. Smirnov[24], J.R. Smith[7], Y. Soloviev[24], J. Spiekermann[8], H. Spitzer[13], R. Starosta[1], M. Steenbock[13], P. Steffen[11], R. Steinberg[2], B. Stella[32], K. Stephens[22], J. Stier[11], J. Stiewe[15], U. Stösslein[35], J. Strachota[30], U. Straumann[37], W. Struczinski[2], J.P. Sutton[3], S. Tapprogge[15], R.E. Taylor[38,27], V. Tchernyshov[23], C. Thiebaux[28], G. Thompson[20], P. Truöl[37], J. Turnau[6], J. Tutas[14], P. Uelkes[2], A. Usik[24], S. Valkár[31], A. Valkárová[31], C. Vallée[25], P. Van Esch[4], P. Van Mechelen[4], A. Vartapetian[11,39], Y. Vazdik[24], M. Vecko[30], P. Verrecchia[9], G. Villet[9], K. Wacker[8], A. Wagener[2], M. Wagener[33], I.W. Walker[18], A. Walther[8], G. Weber[13], M. Weber[11], D. Wegener[8], A. Wegner[11], H.P. Wellisch[26], L.R. West[3], S. Willard[7], M. Winde[35], G.-G. Winter[11], A.E. Wright[22], E. Wünsch[11], N. Wulff[11], T.P. Yiou[29], J. Žáček[31], D. Zarbock[12], Z. Zhang[27], A. Zhokin[23], M. Zimmer[11], W. Zimmermann[11], F. Zomer[27] and K. Zuber[15],

[1] *I. Physikalisches Institut der RWTH, Aachen, Germany*[a]
[2] *III. Physikalisches Institut der RWTH, Aachen, Germany*[a]
[3] *School of Physics and Space Research, University of Birmingham, Birmingham, UK*[b]
[4] *Inter-University Institute for High Energies ULB-VUB, Brussels; Universitaire Instelling Antwerpen, Wilrijk, Belgium*[c]
[5] *Rutherford Appleton Laboratory, Chilton, Didcot, UK*[b]
[6] *Institute for Nuclear Physics, Cracow, Poland*[d]
[7] *Physics Department and IIRPA, University of California, Davis, California, USA*[e]
[8] *Institut für Physik, Universität Dortmund, Dortmund, Germany*[a]
[9] *CEA, DSM/DAPNIA, CE-Saclay, Gif-sur-Yvette, France*
[10] *Department of Physics and Astronomy, University of Glasgow, Glasgow, UK*[b]
[11] *DESY, Hamburg, Germany*[a]
[12] *I. Institut für Experimentalphysik, Universität Hamburg, Hamburg, Germany*[a]
[13] *II. Institut für Experimentalphysik, Universität Hamburg, Hamburg, Germany*[a]
[14] *Physikalisches Institut, Universität Heidelberg, Heidelberg, Germany*[a]
[15] *Institut für Hochenergiephysik, Universität Heidelberg, Heidelberg, Germany*[a]
[16] *Institut für Reine und Angewandte Kernphysik, Universität Kiel, Kiel, Germany*[a]
[17] *Institute of Experimental Physics, Slovak Academy of Sciences, Košice, Slovak Republic*[f]
[18] *School of Physics and Materials, University of Lancaster, Lancaster, UK*[b]
[19] *Department of Physics, University of Liverpool, Liverpool, UK*[b]
[20] *Queen Mary and Westfield College, London, UK*[b]
[21] *Physics Department, University of Lund, Lund, Sweden*[g]
[22] *Physics Department, University of Manchester, Manchester, UK*[b]
[23] *Institute for Theoretical and Experimental Physics, Moscow, Russia*

[24] *Lebedev Physical Institute, Moscow, Russia*[f]
[25] *CPPM, Université d'Aix-Marseille II, IN2P3-CNRS, Marseille, France*
[26] *Max-Planck-Institut für Physik, München, Germany*[a]
[27] *LAL, Université de Paris-Sud, IN2P3-CNRS, Orsay, France*
[28] *LPNHE, Ecole Polytechnique, IN2P3-CNRS, Palaiseau, France*
[29] *LPNHE, Universités Paris VI and VII, IN2P3-CNRS, Paris, France*
[30] *Institute of Physics, Czech Academy of Sciences, Praha, Czech Republic*[f,h]
[31] *Nuclear Center, Charles University, Praha, Czech Republic*[f,h]
[32] *INFN Roma and Dipartimento di Fisica, Universita "La Sapienza", Roma, Italy*
[33] *Paul Scherrer Institut, Villigen, Switzerland*
[34] *Fachbereich Physik, Bergische Universität Gesamthochschule Wuppertal, Wuppertal, Germany*[a]
[35] *DESY, Institut für Hochenergiephysik, Zeuthen, Germany*[a]
[36] *Institut für Teilchenphysik, ETH, Zürich, Switzerland*[i]
[37] *Physik-Institut der Universität Zürich, Zürich, Switzerland*[i]
[38] *Stanford Linear Accelerator Center, Stanford California, USA*
[39] *Visitor from Yerevan Phys.Inst., Armenia*
[40] *Visitor from Institute of Physics, Azerbaijan Academy of Sciences, Baku, Azerbaijan* [a]

*Supported by the Bundesministerium für Forschung und Technologie, FRG under contract numbers 6AC17P, 6AC47P, 6DO57I, 6HH17P, 6HH27I, 6HD17I, 6HD27I, 6KI17P, 6MP17I, and 6WT87P*
[b] *Supported by the UK Particle Physics and Astronomy Research Council, and formerly by the UK Science and Engineering Research Council*
[c] *Supported by FNRS-NFWO, IISN-IIKW*
[d] *Supported by the Polish State Committee for Scientific Research, grant No. 204209101*
[e] *Supported in part by USDOE grant DE F603 91ER40674*
[f] *Supported by the Deutsche Forschungsgemeinschaft*
[g] *Supported by the Swedish Natural Science Research Council*
[h] *Supported by GA ČR, grant no. 202/93/2423 and by GA AV ČR, grant no. 19095*
[i] *Supported by the Swiss National Science Foundation*

# 1 Introduction

The measurement of the inclusive deep inelastic lepton-proton scattering cross section has been of great importance for the understanding of quark-gluon substructure of the proton [1]. Experiments at HERA extend the previously accessible kinematic range up to very large squared momentum transfers, $Q^2 > 5{\cdot}10^4$ GeV$^2$, and to very small values of Bjorken $x < 10^{-4}$. In 1993 the first observation was reported of a rise of the proton structure function $F_2(x,Q^2)$ at low $x < 10^{-2}$ with decreasing $x$ [2, 3]. Such a behaviour is qualitatively expected in the double leading log limit of Quantum Chromodynamics [4]. It is, however, not clarified whether the linear QCD evolution equations, as the conventional DGLAP evolution [5] in $\ln Q^2$ and/or the BFKL evolution [6] in $\ln(1/x)$, describe the rise of $F_2$ or whether there is a significant effect due to nonlinear parton recombination [7]. The quantitative investigation of the quark-gluon interaction dynamics at low $x$ is one of the major challenges at HERA. It requires high precision for the $F_2$ measurement and an investigation of the hadronic final state.

The structure function $F_2(x,Q^2)$ is derived from the inclusive lepton-proton scattering cross section. It depends on the squared four momentum transfer $Q^2$ and the scaling variable $x$. These variables are related to the inelasticity parameter $y$ and to the total squared centre of mass energy of the collision $s$ as $Q^2 = xys$ with $s = 4E_eE_p$. In 1993 the incident electron energy was $E_e = 26.7$ GeV and the proton energy was $E_p = 820$ GeV. A salient feature of the HERA collider experiments is the possibility of measuring not only the scattered electron but also the complete hadronic final state, apart from losses near the beam pipe. This means that the kinematical variables $x$, $y$ and $Q^2$ can be determined with complementary methods which experimentally are sensitive to different systematic effects. The comparison of the results obtained with different methods improves the accuracy of the $F_2$ measurement. A convenient combination of the results ensures maximum coverage of the available kinematic range.

In this paper an analysis is presented of inclusive deep inelastic scattering (DIS) data taken by the H1 collaboration in 1993 with an integrated luminosity of 0.271 pb$^{-1}$. During 1993, its second year of operation, HERA delivered an integrated luminosity of an order of magnitude larger than in 1992. With these increased statistics the accessible kinematic range has been extended and the behaviour of $F_2$ has been investigated at a new level of precision. A similar measurement was published recently by the ZEUS collaboration [8].

The structure function results presented here are more precise with a typical systematic error of 10% than the previous H1 data. The following new information is provided: i) analyzing data with shifted vertex position along the proton beam line, the first DIS data for $Q^2$ between 5 and 10 GeV$^2$ at $x \leq 0.001$ is obtained in a region where new effects could be observed; ii) the analysis extends down to $y \simeq 0.01$ which allows for the first time to approach with HERA data the region of the fixed target lepton-proton scattering experiments; iii) due to the increased statistics, the first precise H1 measurement of $F_2$ beyond $Q^2 \simeq 100$ GeV$^2$ can be presented. An analysis is presented of the dependence of $F_2$ on the effective mass $W$ of the virtual photon-proton system which, due to the very large energy $s$ at HERA, can be performed in a new kinematic range.

A first measurement is given of the diffractive contribution to $F_2$ by analyzing the subsample of about 6% of the DIS events which exhibit no activity in the forward detector region where “forward” denotes the direction of the proton beam. Characteristics of the events with a large pseudo-rapidity gap in the hadron flow close to the proton remnant direction have already been studied at HERA [9, 10].

The paper is organized as follows. After a brief introduction to the kinematics of the inclusive $ep$ scattering process (sec.2), the H1 apparatus and the 1993 data taking (sec.3) are described. Then the event selection including the background rejection (sec.4) and the detector response (sec.5) are discussed. Section 6 describes the $F_2$ analyses and discusses the results. In section 7 the diffractive data analysis is presented. The paper is summarized in section 8.

## 2 Kinematics and Analysis Methods

The kinematic variables of the inclusive scattering process $ep \rightarrow eX$ can be reconstructed in different ways using measured quantities from the hadronic final state and from the scattered electron. The choice of the reconstruction method for $Q^2$ and $y$ determines the size of systematic errors, acceptance and radiative corrections. The basic formulae for $Q^2$ and $y$ used in the different methods are summarized below, $x$ being obtained from $Q^2 = xys$. For the electron (“E”) method

$$y_e = 1 - \frac{E'_e}{E_e} \sin^2 \frac{\theta_e}{2} \qquad Q^2_e = \frac{E'^2_e \sin^2 \theta_e}{1 - y_e}, \tag{1}$$

where the electron polar angle is defined with respect to the incident proton beam direction ($+z$ axis). The resolution in $Q^2_e$ is 4% while the precision of the $y_e$ measurement degrades as $1/y_e$. Thus the electron method cannot be used for $y_e \leq 0.05$. In the low $y$ region it is, however, possible to use the hadronic methods for which it is convenient to define the following variables

$$\Sigma = \sum_i (E_i - p_{z,i}) \qquad (p^h_T)^2 = (\sum_i p_{x,i})^2 + (\sum_i p_{y,i})^2. \tag{2}$$

Here $E, p_x, p_y, p_z$ are the four-momentum vector components of each particle and the summations extend over all hadronic final state particles. The standard definitions for $y_h$ [11] and $\theta_h$ are

$$y_h = \frac{\Sigma}{2E_e} \qquad \tan \frac{\theta_h}{2} = \frac{\Sigma}{p^h_T}. \tag{3}$$

The combination of $y_h$ and $Q^2_e$ defines the mixed method [12] which is well suited for medium and low $y$ measurements. It was used in the previous H1 analysis of $F_2$ [2]. The double-angle (“DA”) method [13] makes use only of $\theta_e$ and $\theta_h$ with

$$y_{DA} = \frac{\tan(\theta_h/2)}{\tan(\theta_e/2) + \tan(\theta_h/2)} \qquad Q^2_{DA} = 4E^2_e \frac{\cot(\theta_e/2)}{\tan(\theta_e/2) + \tan(\theta_h/2)}. \tag{4}$$

The method is rather insensitive to the absolute energy calibration of the detector. It has good resolution at large $Q^2$ where the jet energies are high but the resolution deteriorates for $x \leq 0.001$. The formulae for the $\Sigma$ method [14] are constructed requiring $Q^2$ and $y$ to be independent of the incident electron energy. Replacing in eq.3 the quantity $2E_e$ by $\Sigma + E_e'(1 - \cos\theta_e)$, as allowed by the conservation of the total $E - P_z$ of the event[1], and $(1 - y_e)$ by $(1 - y_\Sigma)$ in eq.1, one obtains:

$$y_\Sigma = \frac{\Sigma}{\Sigma + E_e'(1 - \cos\theta_e)} \qquad \text{and} \qquad Q^2_\Sigma = \frac{E_e'^2 \sin^2\theta_e}{1 - y_\Sigma}. \tag{5}$$

For non radiative events $y_h$ and $y_\Sigma$ are equivalent at low $y$. However, the modified quantity $y_\Sigma$ is less sensitive than $y_h$ to the imprecise hadronic measurement at high $y$ since the $\Sigma$ term dominates the $E - P_z$ of the event. Therefore the $\Sigma$ method can be applied over the full kinematic range considered in this paper.

All these methods were utilized to measure $F_2$. Resolution values of the $x$ and $Q^2$ variables reconstructed with the H1 detector are discussed for all methods at the end of section 5.

# 3 The H1 Detector and the Data Taking in 1993

## 3.1 The H1 Detector

The H1 detector [15] is a nearly hermetic multi-purpose apparatus built to investigate the inelastic high-energy interactions of electrons and protons at HERA. The structure function measurement relies essentially on the inner tracking chamber system, on the backward electromagnetic calorimeter, and on the liquid argon calorimeters which are described briefly below. For the luminosity measurement see section 3.2.

The tracking system includes the central tracking chambers, the forward tracker modules and a backward proportional chamber. These chambers are placed around the beam pipe at $z$ positions between $-1.5$ m and 2.5 m. A superconducting solenoid surrounding both the tracking system and the liquid argon calorimeter provides a uniform magnetic field of 1.15 T.

The central jet chamber (CJC) consists of two concentric drift chambers covering a polar angle of 15° to 165°. Tracks crossing the CJC are measured with a transverse momentum resolution of $\delta p_T/p_T < 0.01 \cdot p_T/\mathrm{GeV}$. The CJC is supplemented by two cylindrical drift chambers at radii of 18 and 47 cm, respectively, to determine the $z$ coordinate of the tracks. To each of the $z$ drift chambers a proportional chamber is attached for triggering. The inner one (CIP) was used in addition to estimate residual photoproduction background.

A tracking chamber system made of three identical modules measures charged particles emitted in forward direction (3° to 20°). This forward tracker (FT) is used to determine the vertex for the events which leave no track in the CJC. This extends the analysis into the larger $x$ region.

In the backward region a four plane multiwire proportional chamber (BPC) provides a space point for charged particles with a spatial resolution of about 1.5 mm in the transverse plane. The polar angle acceptance of the BPC ranges from 155° to 174.5°. The reconstructed space point together with the $z$ vertex position defines the polar angle of the scattered electrons.

The BPC is attached to the backward electromagnetic calorimeter (BEMC) which is made of 88 lead/scintillator stacks with a size of $16 \times 16$ cm$^2$ and a depth of 22 radiation lengths,

[1] defined as $(E - P_z) \equiv \sum_j (E_j - p_{z,j})$, the sum extending over all particles $j$ of the event.

corresponding to about one hadronic interaction length. The angular coverage of the BEMC is $155^o < \theta < 176^o$. A 1.5 cm spatial resolution of the lateral shower position is achieved using four photodiodes which detect the wavelength shifted light from each of the scintillator stacks. A scintillator hodoscope (TOF) situated behind the BEMC is used to veto proton-induced background events based on their early time of arrival compared with nominal $ep$ collisions.

Hadronic final state energies and electrons at high $Q^2$ are measured in the highly segmented liquid argon (LAr) calorimeter [16] which covers an angular region between $3^o$ and $155^o$. The LAr calorimeter consists of an electromagnetic (e.m.) section with lead absorber and a hadronic section with stainless steel absorber. The electromagnetic part has a depth between 20 and 30 radiation lengths. The total depth of both calorimeters varies between 4.5 and 8 interaction lengths. The most backward part of the LAr calorimeter is a smaller electromagnetic calorimeter (BBE) [2] which covers the polar angle range from $146^o$ to $155^o$.

The DIS events in the backward region (low $Q^2$ data, $\theta_e > 155^o$) were triggered by an energy cluster in the BEMC ($E > 4$ GeV) which was not vetoed by an out of time signal in the TOF. The high $Q^2$ events were triggered by requiring an e.m. energy cluster with $E > 8$ GeV in the LAr calorimeter. For lower energy thresholds ($> 5$ GeV) the events were also triggered if there was simultaneously a tracking trigger. The trigger efficiency has been determined from the data using the redundancy of the H1 trigger system which relies on calorimetry and tracking. In the region of the final $F_2$ data presented below, i.e. for $E'_e > 10.6$ GeV, the DIS electron trigger efficiency is 100% within the errors.

### 3.2 Data Samples and Luminosity

In 1993 HERA was operated with 84 colliding electron and proton bunches. Sixteen pilot bunches, 6 proton and 10 electron, underwent no $ep$ collision. From these the beam induced background could be estimated. A small part of the data was taken with the nominal interaction position shifted in $z$ by +80 cm in order to reach $Q^2$ values as low as 4 GeV$^2$. The lower $Q^2$ region was covered also by analyzing events which originated from the so-called early "proton satellite" bunch which collided with an electron bunch at $z \simeq +62$ cm. The satellite bunch data amount to $\simeq 3\%$ of the total data. Both event samples with shifted $z$ vertex positions were analyzed independently and the results combined.

To reduce the systematic errors of the $F_2$ measurement, a strict data selection was performed based on the behaviour of the main detector components. In particular, data taken during a period in which there was no magnetic field due to a failure of the coil were excluded ($\simeq 0.15$ pb$^{-1}$). The number of accepted events per luminosity was checked to be constant within statistical errors during the full data taking period.

The luminosity was determined from the measured cross section of the Bethe Heitler reaction $e^-p \rightarrow e^-p\gamma$. The final state electron and the photon are simultaneously detected in small electromagnetic calorimeters (electron and photon “taggers”) close to the beam pipe but at a large distance from the main detector (at $z = -33$ m and $z = -102$ m). This coincidence method gives the best on-line luminosity estimate but it is sensitive to variations in the electron beam optics through the electron tagger acceptance. Therefore the final luminosity determination was performed based on the hard photon bremsstrahlung data using the photon tagger only. The results of these two methods differ by 1.1%. This is well below the overall systematic error which is estimated to be equal to 4.5% [17].

[2] The H1 detector calorimetry and electron detection is thus split into two parts. At large angles, $\theta > 155^o$, the electron is measured in the BEMC and preceding tracking chambers. This defines for the subsequent analysis the low $Q^2$ data sample. Electrons at lower polar angles, $\theta < 155^o$, are detected in the LAr calorimeter and the forward and central chambers defining the high $Q^2$ event sample. For the structure function analysis both data samples were combined but the analyses of the low and high $Q^2$ data required partially specific techniques and cuts. Special attention had to be paid to the BEMC-BBE transition region where the electron energy is often shared between the two calorimeters.

An independent check of the luminosity measurement has been performed by determining the cross section for QED Compton events which leave a scattered electron and a photon in the backward part of the apparatus. For the 1993 data this yields a luminosity value which is 4% lower than the one from the Bethe Heitler cross-section measurement, with a combined systematic and statistical error of 8% [17, 18].

The integrated luminosity for the nominal vertex data used in this analysis is 0.271 $pb^{-1}$, and 2.5 $nb^{-1}$ for the special data taking with the shifted vertex position. The luminosity of the satellite data was obtained from the measured luminosity for the nominal vertex data multiplied by the efficiency corrected event ratio in a kinematic region common to both data sets. The error of that luminosity determination was estimated to be 8.5%.

# 4 Event Selection and Monte Carlo Simulation

## 4.1 Event Selection

The event selection criteria can be divided into four categories: i) electron identification, ii) event vertex requirement, iii) kinematical constraints and iv) background rejection. The latter is discussed in the subsequent section. A summary of the selection cuts is given in tables 1 and 2.

i) For the **electron identification** in the backward region of the H1 detector, for $\theta_e > 155^o$, the most energetic cluster in the BEMC is selected. Its center of gravity is required to be at a radial distance (“CLBP”) smaller than 4 cm from a reconstructed BPC point. The lateral size (“ECRA”) of that cluster, calculated as the energy weighted radial distance of the cells from the cluster centre, has to be smaller than 4 cm. In the LAr calorimeter, for $\theta_e < 155^o$, several electron identification algorithms have been developed which have the common feature of demanding significant electromagnetic energy deposited in a confined region (table 2). The E method required an isolated electromagnetic cluster with $\epsilon_4 > 50\%$ and a minimum relative energy (“$\epsilon_3$”) of 3% in the first e.m. layer of the LAr calorimeter which amounts to about three radiation lengths [19]. Here $\epsilon_4$ is the energy sum over the four most energetic cells of a cluster divided by its total energy. For the $\Sigma$ analysis an e.m. cluster with $\epsilon_4 > 65\%$ was considered to be an electron.

In the transition region between the BEMC and the BBE, for $\theta_e$ about $155^o \pm 2^o$, a dedicated E and DA analysis was performed in which the electron finding was based on topological criteria only, i.e. without making use of $E'_e$ explicitely. For the $\Sigma$ analysis a fiducial cut was applied if the electron cluster was shared between the BEMC and the BBE. Because of the smearing of the $z$ vertex position there is no loss in the $x, Q^2$ coverage, only a reduction in statistics.

The electron identification efficiency for DIS events, as determined from Monte Carlo simulation, is better than 97% apart from the BEMC-BBE transition region where it amounts to approximately 92%. The scattered electron is considered to be the cluster with the largest energy. If two electron candidates are present in the same event, the one of highest energy is selected. This introduces a misidentification probability of at most 3% [20] at the smallest electron energy which was considered in the systematic error calculation.

ii) The **event vertex** position is needed for a precise determination of the kinematics. It was defined by at least one well measured track in the CJC or in the FT crossing the beam axis. The $z$ vertex region for the events with interactions in the nominal time intervals was $z = (-5 \pm 30)$ cm. For the satellite bunch data the requirement was $z = (+62 \pm 20)$ cm and for the shifted vertex data $z = (+75 \pm 25)$ cm. The vertex reconstruction efficiency, as determined from the data, is 97% at $y > 0.2$. It decreases smoothly to an average of 73% in the lowest $y$ bins ($y \simeq 0.01$) since with decreasing $y$ the hadronic particles are emitted at smaller polar angles. The vertex reconstruction efficiency for the shifted vertex data is only about 10% lower than for the nominal vertex data since $z \sim 75$ cm is still inside the central tracking system.

iii) The basic **kinematic constraints** were a maximum electron scattering angle of $173^o$ ($175.2^o$ for the shifted vertex data) and minimum energy requirements in order to ensure high trigger efficiency and a small photoproduction background. For the high $Q^2$ data, the minimum energy requirement was replaced by introducing a cut rejecting events at high $y_e$. At low $Q^2$, the minimum electron candidate energy $E'_e$ was 10.6 GeV for the E analysis and 8 GeV for the $\Sigma$ analysis. That difference is an example of selection criteria differences arising due to the combination of several complete and independent analyses. The $\Sigma$ analysis also required a total missing transverse momentum smaller than 15 GeV and a total $E-P_z$ between 30 and 75 GeV. In analyzing the BEMC-BBE transition region alternative requirements, like $log(y_{DA}/y_h) < 0.5$, were introduced avoiding the use of $E'_e$ in the selection. The latter conditions rejected photoproduction background events and DIS events with an energetic photon radiated along the electron beam direction, thus reducing the radiative corrections to $F_2$.

| | E method | $\Sigma$ method | DA method |
|---|---|---|---|
| $\theta_e/^o$ | $< 173$ | $< 173$ | $< 173$ |
| $E'_e$/GeV | $> 10.6$ | $> 8$ | $> 10.6$ |
| $z_{vertex}$/cm | $-5\pm$ 30 | $-5\pm$ 30 | $-5\pm$ 30 |
| ECRA/cm | $< 4$ | $< 4$ | $< 4$ |
| CLBP/cm | $< 4$ | $< 4$ | $< 4$ |
| $(E-P_z)$/GeV | – | $< 75$, $> 30$ | – |
| $\theta_h/^o$ | – | – | $> 40$ |
| selected events | 22500 | 24100 | 22200 |

Table 1: *Summary of event selection criteria for the low $Q^2$ data, i.e. for $\theta_e > 155^o$. For the definitions of abbreviations see the text.*

| | E method | $\Sigma$ method | DA method |
|---|---|---|---|
| $\theta_e/^o$ | $< 150$ | $< 155$ | $< 150$ |
| $y_e$ | $< 0.6$ | $< 0.8$ | $< 0.8$ |
| $z_{vertex}/$cm | $-5\pm 30$ | $-5\pm 30$ | $\geq 1$ track |
| electron ident. | $\epsilon_4 > 50\%$ | $\epsilon_4 > 65\%$ | min $R_{cl}$ |
| | $\epsilon_3 > 3\%$ | $E'_e > 8\ GeV$ | $E'_e > 8\ GeV$ |
| $p_T^h/p_T^e$ | $> 0.2$ | – | – |
| $(E - P_z)/$GeV | – | $> 30$ | $> 30$ |
| missing $p_T/$GeV | – | $< 15$ | $< 10 - 25$ |
| selected events | 880 | 1100 | 1038 |

Table 2: *Summary of event selection criteria for the high $Q^2$ data. Further topological requirements based on the LAr calorimeter were imposed to reject background. The quantity $R_{cl}$ in the DA method denotes the average over calorimeter cell energies of the polar angle distance between a cell and the cluster centre which is a measure of the lateral extension of the calorimeter energy deposition. For further definitions see the text.*

## 4.2 Background Rejection

### 4.2.1 Non-*ep* Background

At low $Q^2$ the main sources of non-*ep* background are due to proton beam interactions with residual gas and beam line elements upstream of the H1 detector. Beam wall events in the interaction region were rejected by requiring the reconstructed vertex to be centered on the beam axis. An efficient reduction of the remaining background is provided by the minimum energy and the vertex requirements discussed above. In addition two algorithms have been developed based on the central tracking information. An event was rejected either if the fraction of tracks which did not point to the reconstructed vertex was too high, or if at least two tracks pointed to the backward direction and crossed the beam line at $z < -50$ cm. Both algorithms gave similar results. They introduced a loss of a few per cent of genuine DIS events which was controlled by eye scanning and corrected for. From the study of pilot bunches, the residual background was estimated to represent less than 1% of the total number of selected events.

At high $Q^2$ the main background is due to muons travelling off axis parallel to the proton beam. These are produced by proton beam halo interactions and occasionally generate an electromagnetic shower in the LAr calorimeter. Requiring a reconstructed vertex rejects most of them. Further rejection was obtained by requiring more than 10 GeV deposited outside the calorimeter region which contains the electron candidate cluster. Residual cosmic ray event candidates were rejected with similar topological requirements. The remaining background was estimated to be smaller than 2% by a visual event scan and also by analyzing the pilot bunch data.

### 4.2.2 Photoproduction Background

The only significant background to DIS from *ep* interactions is due to photoproduction events at $Q^2 \simeq 0$ where the scattered electron escapes the detector along the beam pipe but in which an energy cluster from the hadronic final state fakes an electron. About 10% of these events are identified as photoproduction background if the scattered electron is found in the electron tagger.

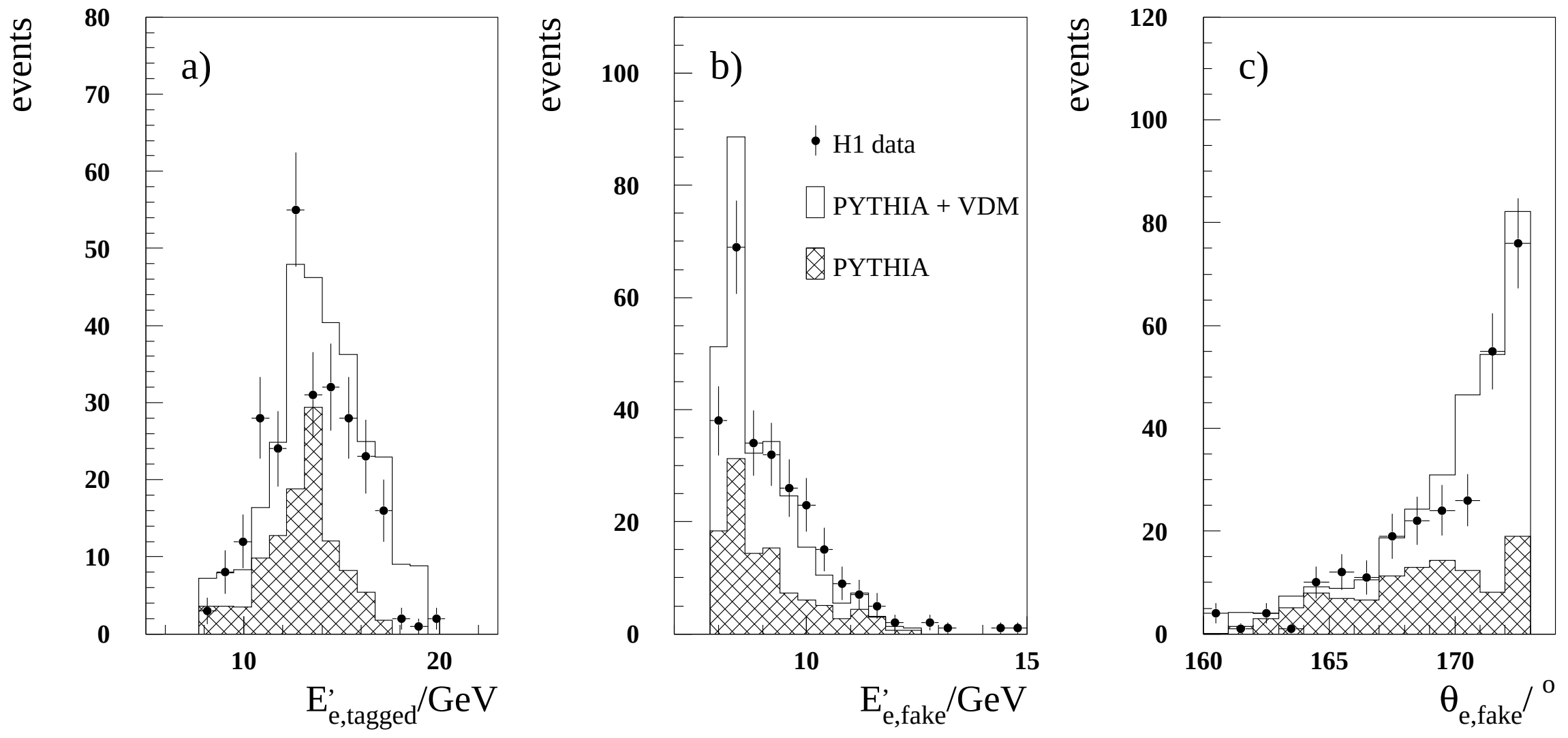

Figure 1: *Distributions for tagged photoproduction events: a) Energy distribution of the scattered electron energy in the electron tagger, b) and c) energy and angle distributions of the fake electron candidate in the BEMC. The statistical errors of the Monte Carlo simulation are of the same order of magnitude as the experimental ones.*

To estimate this background, photoproduction events were simulated corresponding to the luminosity of the data. The "soft" vector meson contribution was simulated using the RAYVDM [21] program, and the "hard" scattering part using the PYTHIA [22] program. For hard scattering interactions, direct and resolved processes and the production of heavy quarks were included. The relative contributions of both were adjusted to agree with the total photoproduction cross section analysis [23]. The simulation both in shape and normalization [24, 25] is in good agreement with the tagged photoproduction data. This is illustrated in fig.1 showing the scattered electron energy distribution in the tagger and the energy and angle distributions of the fake electron detected in the BEMC. The photoproduction background was subtracted statistically bin by bin. The highest contamination is in the lowest $Q^2, x$ bin and amounts to $(9 \pm 4)\%$ for $E_e' > 10.6$ GeV. Only three bins have a contamination larger than 3% [26].

An independent analysis [27] made use of the tracking information to identify that part of the photoproduction background which originates from photons in a low $Q^2$ interaction. These can mimic a DIS electron due to conversion in the CJC end flange. Such events were rejected by requiring a hit in the innermost tracking chamber which is crossed by the particle before reaching the CJC. The remaining background from charged pions and conversions in the beampipe was estimated based on a comparison of the tagged photoproduction events and the Monte Carlo simulation which allowed to subtract this background part statistically. This background estimation gave consistent results with the Monte Carlo method described above.

### 4.3 Monte Carlo Simulation

More than half a million Monte Carlo events were generated using DJANGO [28] and different quark distribution parametrizations, corresponding to an integrated luminosity of approximately 1.5 $pb^{-1}$. The program is based on HERACLES [29] for the electroweak interaction and on LEPTO [30] to simulate the hadronic final state. HERACLES includes first order radiative corrections, the simulation of real bremsstrahlung photons and the longitudinal structure function. The acceptance corrections were performed using the MRSH parametrization [31], which is constrained to the HERA $F_2$ results of 1992. LEPTO uses the colour dipole model (CDM) as implemented in ARIADNE [32] which is in good agreement with data on the energy flow and other characteristics of the final state as measured by H1 [33] and ZEUS [34]. For the estimation of systematic errors connected with the topology of the hadronic final state, the HERWIG model [35] was used in a dedicated analysis. Based on the GEANT program [36] the detector response was simulated in detail. After this step the Monte Carlo events were subject to the same reconstruction and analysis chain as the real data.

## 5 Calibration and Kinematical Distributions

The structure function measurement requires an understanding of the reconstruction of the energies and angles of the scattered electron and of the hadronic final state because all are used to define the kinematics of the interaction. Figs.2 and 3 display the distributions of the reconstructed energy $E_e'$ and angle $\theta_e$ of the scattered electron and of the hadronic variables $y_\Sigma$ and $\theta_h$. Fig.2 shows the distributions for the high statistics, low $Q^2$ event sample with the electron measured in the BEMC. The small contamination due to photoproduction background is also displayed. The Monte Carlo distributions are normalized to the luminosity measurement. Agreement between data and Monte Carlo simulation was obtained after an iteration of the structure function parametrization (see below). The remaining small discrepancies, as visible for example in the distribution of $y_\Sigma$ around 0.3, have negligible influence on the acceptance calculation for $F_2$. Fig.3 shows the same distributions for the smaller sample at large $Q^2$ with the electron detected in the LAr calorimeter. Data and simulation agree well.

A crucial part of the $F_2$ analysis is the absolute energy calibration. The determination of the energy scale of the BEMC is based on the observed shape of the kinematic peak, see fig.2a, for energies between 22 and 28 GeV. Taking into account stack to stack variations of the response and correcting for dead material in front of the BEMC as well as for energy losses between the stacks, the absolute scale of the $E_e'$ measurement in the BEMC was determined with 1.7% systematic and 1% statistical accuracy [37]. The result was cross checked with the double angle method to reconstruct the electron energy and agreement was found to within the quoted uncertainty. For the high $Q^2$ data the double angle method was used to refine the energy scale determined by test beam measurements [38]. The resulting systematic uncertainties in $E_e'$ are smaller than 5% for $\theta_e$ between $150^o$ and $155^o$ (BBE region) and 3% for electrons in the barrel part of the calorimeter ($\theta_e \leq 150^o$).

The hadronic energy scale in the liquid argon calorimeter is presently known to 6% as determined from studies of the transverse momentum balance of DIS events. Test beam data of pions between 3.7 GeV and 205 GeV showed agreement on the 3% level with the

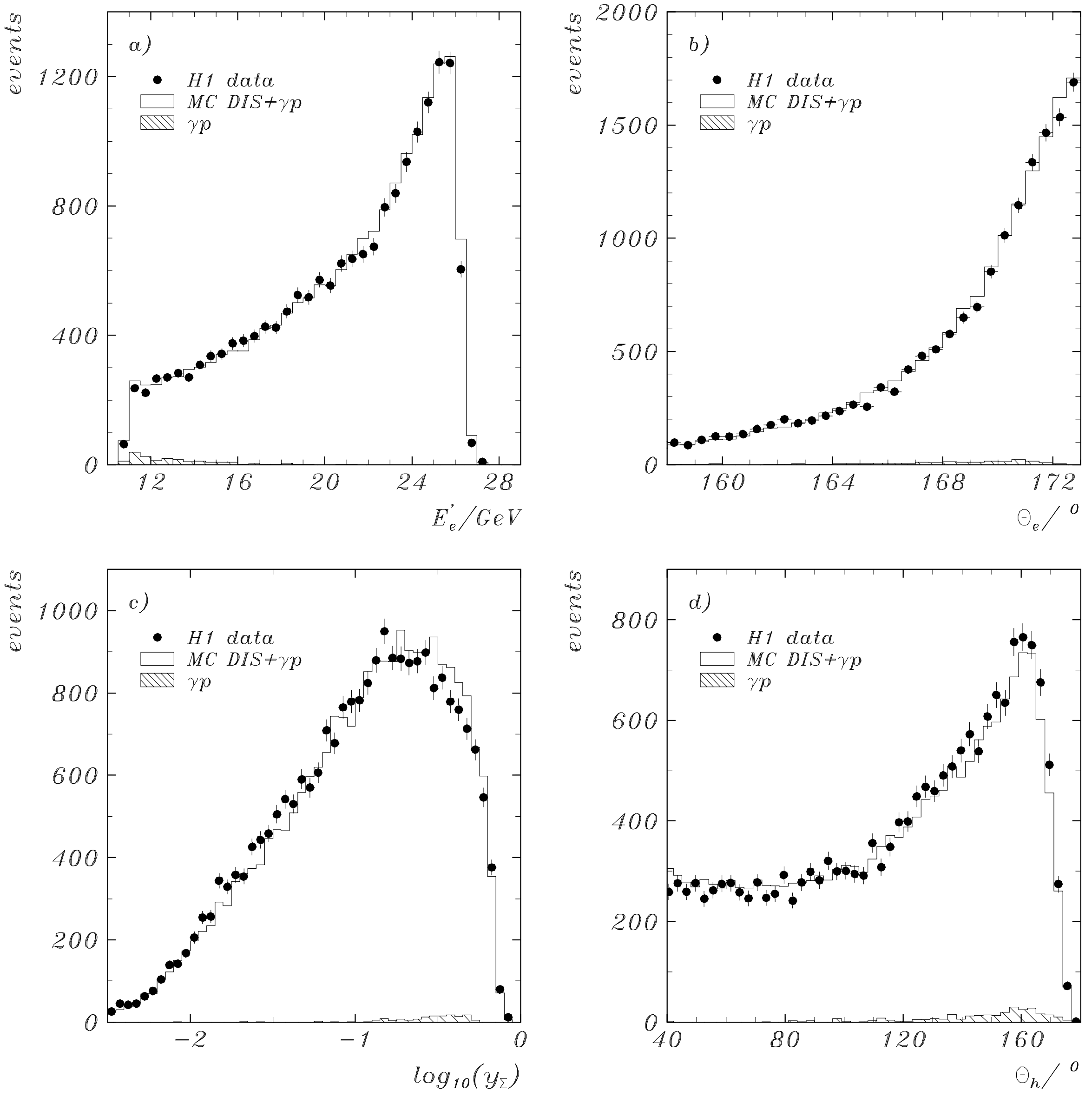

Figure 2: *Distributions of the kinematical quantities $E'_e, \theta_e, y_\Sigma$ and $\theta_h$ for the low $Q^2$ data (closed circles) and the Monte Carlo simulation (open histogram). The Monte Carlo calculation is normalized to the luminosity. The hatched histogram is the estimation of the background due to photoproduction processes.*

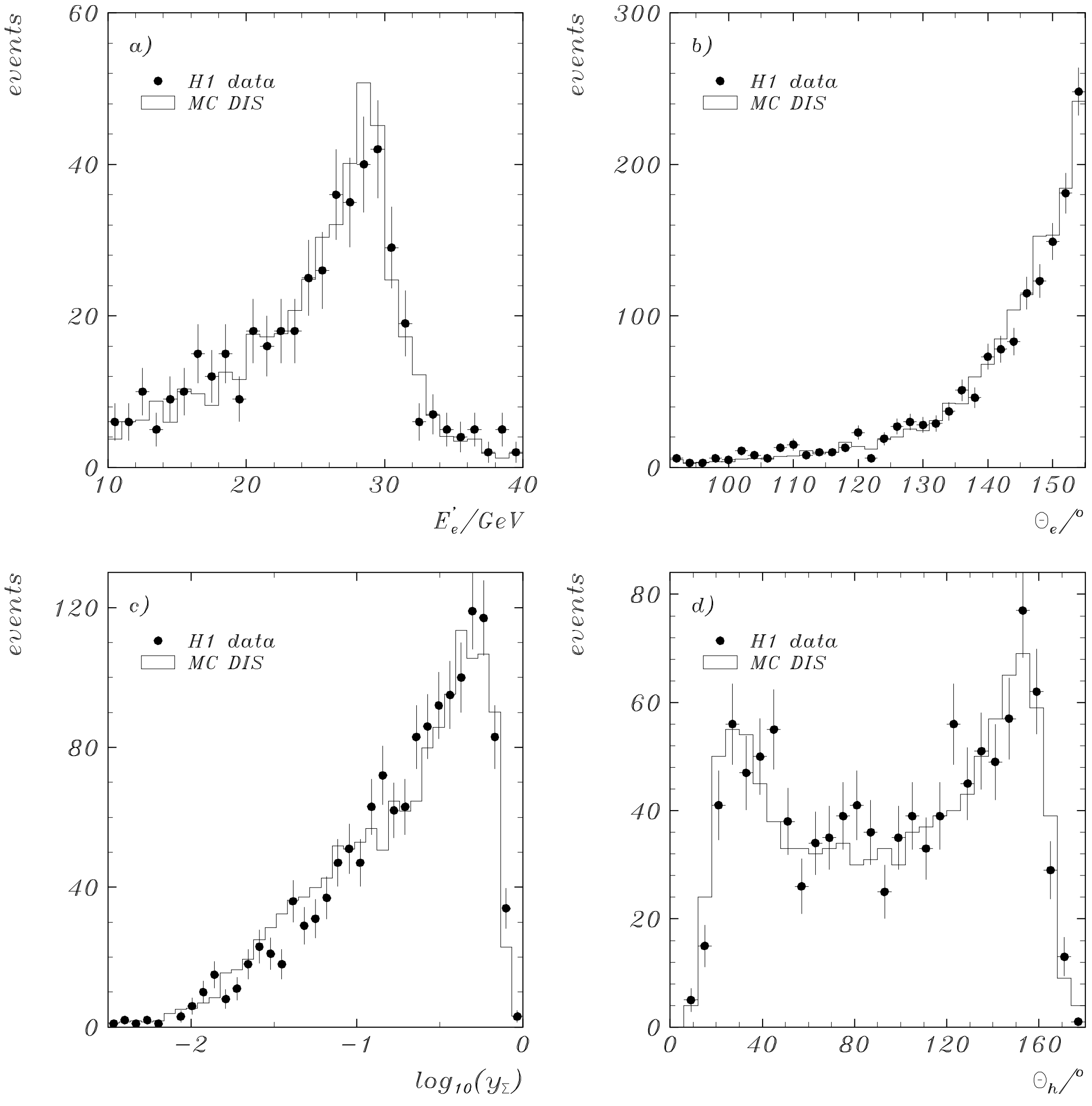

Figure 3: *Distributions of the kinematical quantities* $E_e', \theta_e, y_\Sigma$ *and* $\theta_h$ *for the high* $Q^2$ data ($\theta_e < 155^o$) *(closed circles) and the Monte Carlo simulation (open histogram) using the MRSH parametrization [31] as input structure function. For the energy distribution* $Q_e^2 > 250$ GeV$^2$ *is required. For the* $\theta_h$ *distribution a cut at* $y_{DA} > 0.05$ *is applied to exclude the region where* $F_2^{DA}$ *is not measured.*

Monte Carlo description [39]. The calibration parameters were determined from Monte Carlo simulated jet data [15, 40]. The calorimetric energy due to charged particles in the central detector region was replaced by the more precise momentum measurement of the CJC which renders $y_h$ sensitive to charged hadrons of low momentum [41]. This improved the resolution of $y_h$ by about 40% at low $Q^2$ and high $y$.

An extension to higher $y_h$ values from 0.4, as in the previous H1 analysis [2], to 0.7 was achieved by using the $\Sigma$ method. The behaviour of $y_\Sigma$ can be studied with the high $y$ data by means of the ratio $y_\Sigma/y_e$ (fig.4a) using the excellent resolution of $y_e$ ($\simeq 4\%$) as a reference. As can be seen, both the mean value and the measured resolution (13%) are well reproduced by the Monte Carlo simulation. Good agreement is also achieved in the $p_T^h/p_T^e$ distribution at high $y$ (fig.4b) although the hadronic final state particles are of rather low energy and on average are emitted into the backward direction. The resolution of $p_T^h$ ($\sim 35\%$) is wider than the $p_T^e$ resolution. The coverage of the full $y$ range necessitates an understanding of the effect of the calorimeter noise on $F_2$. For low $y$ this is sensitive in particular in the backward part of the detector where any energy deposition of 300 MeV generates an additive contribution to $y$ of about 0.01. Fig.2c suggests that $y_\Sigma$ values below 0.01 can be reconstructed. In fig.4c the $y_\Sigma/y_{gen}$ distribution is shown for $0.005 < y_\Sigma < 0.015$ including and excluding the calorimeter noise contribution for the reconstruction of $y$. The distributions have a maximum near 1 and the resolution obtained is 27% when the noise is included. Note that for low $y$ the quantity $y_e$ cannot be considered as a measure for the true $y$ since its resolution function is distorted. The contribution of the noise which has been obtained from experimental data is apparently small. The tail of the distribution at large $y_\Sigma/y_{gen}$ is related to the finite granularity of the calorimeter [42]. Good agreement between data and Monte Carlo in the low $y$ region has been achieved. This can be seen, for example, in the distribution of the ratio $p_T^h/p_T^e$, fig.4d. Therefore, the $F_2$ measurement could be extended to $y = 0.01$ and, for the first time, the $x$ region measured at HERA reaches the domain of larger $x$ values covered in fixed target experiments [43, 44].

The electron angle $\theta_e$ has been determined for the lower $Q^2$ data from the vertex position, reconstructed with the central and forward chambers, and the hit in the BPC closest to the position of the electromagnetic cluster in the BEMC. The precision of the $z$ vertex measurement for most of the events is about 1 cm. The $\theta_e$ measurement, in part of the kinematic range, could be validated and cross calibrated with the innermost tracking chamber and the cluster position [27]. Potential shifts of the $\theta_e$ values are estimated to be smaller than 2 mrad and a resolution of 2.5 mrad is achieved. For the high $Q^2$ data, the vertex position and the cluster centre define $\theta_e$ within 5 mrad accuracy and 7 mrad resolution as cross checked with the central jet chamber. For the intermediate $Q^2$ analysis, the $\theta_e$ measurement is made with the CJC and the $z$ drift chambers.

The hadronic angle $\theta_h$ is reconstructed according to eq.3 from the energy deposited in the calorimeter cells. The $\theta_h$ Monte Carlo distribution depends crucially on the shape of the input structure function. Agreement between data and Monte Carlo was obtained after iteration of the input by starting with the MRSH distribution, extracting and fitting $F_2$, and reweighting the Monte Carlo event distributions by the ratio of the fitted $F_2$ to the MRSH parametrization. The weights applied deviate from 1 by less than 20%. While this procedure significantly improved the description of the event distributions shown in fig.2, it had less than 5% influence on the structure function derivation because the dependence of the $F_2$ analysis on the assumed shape of $F_2$ is due only to smearing corrections which are less than 30% in the kinematic range and for the binning used here.

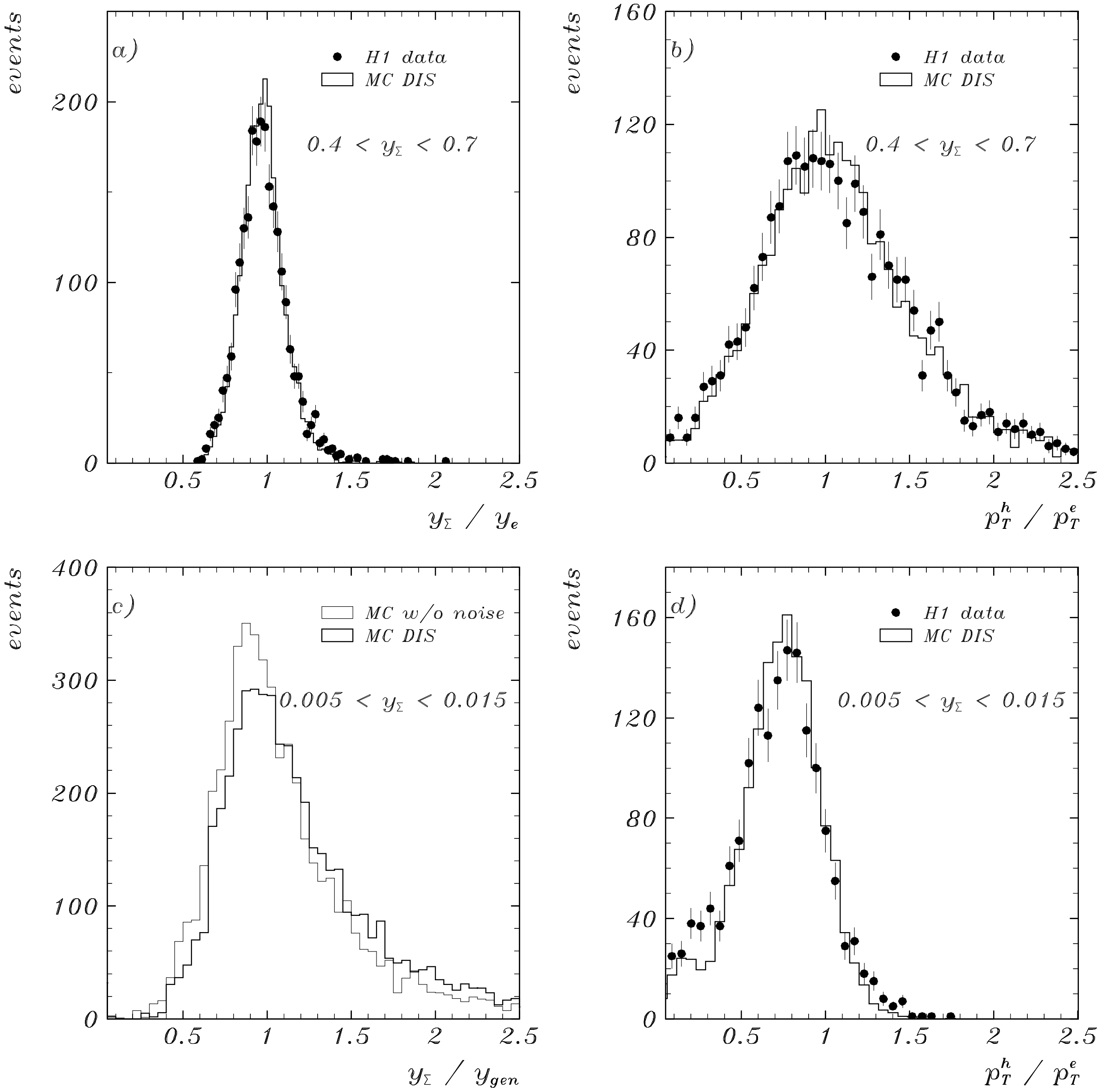

Figure 4: *Experimental and Monte Carlo distributions of* a) $y_\Sigma/y_e$ *and* b) $p_T^h/p_T^e$ *in the high* $y$ *range for* $Q_\Sigma^2 > 10\ GeV^2$. *Monte Carlo distributions of* c) $y_\Sigma/y_{gen}$ *with and without (w/o) the calorimeter noise included and* d) *experimental and Monte Carlo distributions of* $p_T^h/p_T^e$ *at very low* $y$. *The* $p_T$ *ratio distributions at low* $y$ *(d) peak at about 0.8 due to losses of hadrons in the beam pipe both in the data and in the Monte Carlo simulation.*

Detailed studies were performed of the resolution and systematic shifts of the reconstructed $x$ and $Q^2$ values. In table 3 the average resolution values are quoted for the three methods used to determine $F_2$. The $(x, Q^2)$ binning was adapted to the resolution in $x$ and to the statistics in $Q^2$. A rather fine grid is obtained at low $Q^2$ with 8 bins per order of magnitude for $Q^2$ between 7.5 and 133 GeV$^2$. Statistics allowed for four more bins, equidistant in $\log Q^2$, between 133 and 2070 GeV$^2$. Independently of $Q^2$ the binning in $x$ was 6 (4) bins per order of magnitude for $x < (>) 10^{-3}$. The $F_2$ measurement is based on kinematic variables which in the full $(x, Q^2)$ range are reconstructed with systematic shifts smaller than 5% and relative resolutions better than 20%, apart from three edge bins at very low $y$ with a resolution better than 30%.

| | | $x_E$ | $x_\Sigma$ | $x_{DA}$ | $Q^2_E$ | $Q^2_\Sigma$ | $Q^2_{DA}$ |
|---|---|---|---|---|---|---|---|
| $y > 0.15$ | low $Q^2$ | 12 | 13 | 39 | 4 | 11 | 12 |
| | high $Q^2$ | 9 | 10 | 9 | 3 | 9 | 4 |
| $y < 0.15$ | low $Q^2$ | 40 | 19 | 23 | 4 | 7 | 3 |
| | high $Q^2$ | 37 | 16 | 12 | 3 | 5 | 4 |

Table 3: *Relative resolution values in per cent for the three methods used to reconstruct $x$ and $Q^2$.*

# 6 Structure Function Measurement

## 6.1 Comparison of $F_2$ Analyses

The structure function $F_2(x, Q^2)$ was derived after radiative corrections from the one-photon exchange cross section

$$\frac{d^2\sigma}{dxdQ^2} = \frac{2\pi\alpha^2}{Q^4 x}(2 - 2y + \frac{y^2}{1+R})F_2(x, Q^2) \equiv \kappa \cdot F_2 \tag{6}$$

Effects due to $Z$ boson exchange are smaller than 5% and were corrected for at high $Q^2$. The structure function ratio $R = F_2/2xF_1 - 1$ has not been measured yet at HERA and was calculated using the QCD relation [45] and the MRSH structure function parametrization. At lowest $x$ and $Q^2$ the assumed $R$ values are about 0.3, all values being quoted in the $F_2$ data summary table, see below. Compared to the previous H1 analysis [2] the $F_2$ measurement has been extended to lower and higher $Q^2$ (from $8.5 - 60$ GeV$^2$ to $4.5 - 1600$ GeV$^2$) and also to larger $x$.

The determination of the structure function requires the measured event numbers to be converted to the bin averaged cross section based on the Monte Carlo acceptance calculation. The mean acceptance [3] was 0.89. All detector efficiencies were determined from the data utilizing the redundancy of the apparatus. Apart from very small extra corrections all efficiencies are correctly reproduced by the Monte Carlo simulation. The bin averaged cross section was corrected for higher order QED radiative contributions and a bin size correction was performed. This determined the one-photon exchange cross section which according to eq.6 led to the values for $F_2(x, Q^2)$.

[3] The data sample contains a fraction of 6% of large rapidity gap events. With a special rapidity gap Monte Carlo program [46] acceptances were calculated and they agree within statistical errors with those calculated using the standard DIS simulation. Thus only the DIS Monte Carlo simulation has been used for the calculation of the acceptance corrections.

Several methods were used to derive $F_2$, each with different advantages: i) the E method which has the best resolutions on $x$ and $Q^2$ at large $y$ and is independent of the hadron reconstruction, apart from the vertex requirement; ii) the $\Sigma$ method which has small radiative corrections and extends from very low to large $y$ values; iii) the DA method which is less sensitive to the energy scales. A complete analysis [47] has been performed based on the mixed method which agrees very well with the other structure function determinations presented in this paper. The application of different methods was important to check that the systematic errors were correctly evaluated.

The $F_2$ analyses for the shifted vertex and the satellite bunch data were performed with only the E method. The results were found to be in good agreement with each other and with the $F_2$ obtained from the nominal vertex data in the region of overlap. Taking into account correlations between the systematic errors, the two data sets were combined. Thus the first structure function data for $Q^2 \simeq 5$ GeV$^2$ and $x \geq 2 \cdot 10^{-4}$ are presented here.

All structure function values are shown in fig.5. As can be seen the agreement between the analyses is very good. A combination of $F_2^E$ and $F_2^\Sigma$ was chosen for the final result and $F_2^{DA}$ displayed to demonstrate consistency.

### 6.2 Systematic Errors

The systematic errors considered are:

- A potential miscalibration of the electron energy by 1.7% in the BEMC, by 5% in the BBE, and by 3% in the central calorimeter region.

- A 6% scale error for the hadronic energy in the LAr calorimeter, the effect of which is reduced due to the joint consideration of tracks and calorimeter cells for the $\Sigma$ analysis. A 20% scale error was assigned to the energy of the hadronic final state measured in the BEMC. These numbers include uncertainties due to the noise treatment for the LAr calorimeter and the BEMC.

- A shift of up to 2 mrad for the electron polar angle in the BEMC region and of at most 5 mrad in the LAr calorimeter. Errors of the energy and angular resolutions are taken into account in addition to possible shifts of mean values.

- Apart from the electron identification all efficiencies were determined from the data and compared with the Monte Carlo simulation. Agreement between the experimental and the simulated values for the individual efficiencies (trigger, TOF, vertex, CJC track, BPC hit, BEMC and LAr calorimeter cluster reconstruction) was found to be better than 2%. An overall error of 4% was assigned due to the imperfect description of the various efficiencies. A larger error of at most 12% was added to account for the vertex reconstruction efficiency variation at large $x$.

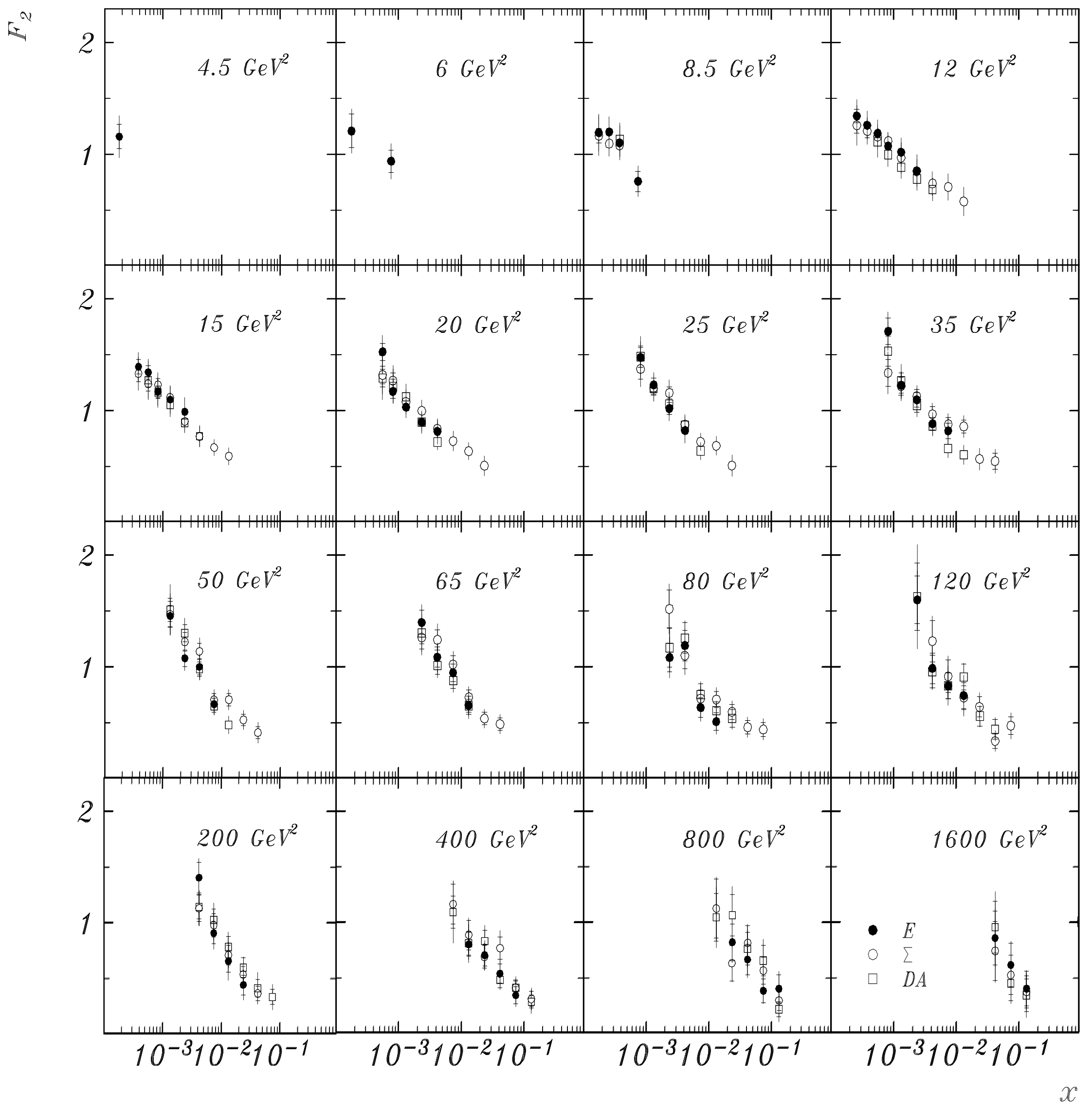

Figure 5: *Measurement of the structure function $F_2(x, Q^2)$ with the electron (closed circles), the $\Sigma$ (open circles) and the double angle method (open squares). The inner error bar is the statistical error. The full error represents the statistical and systematic errors added in quadrature, not taking into account the 4.5% systematic error of the luminosity measurement.*

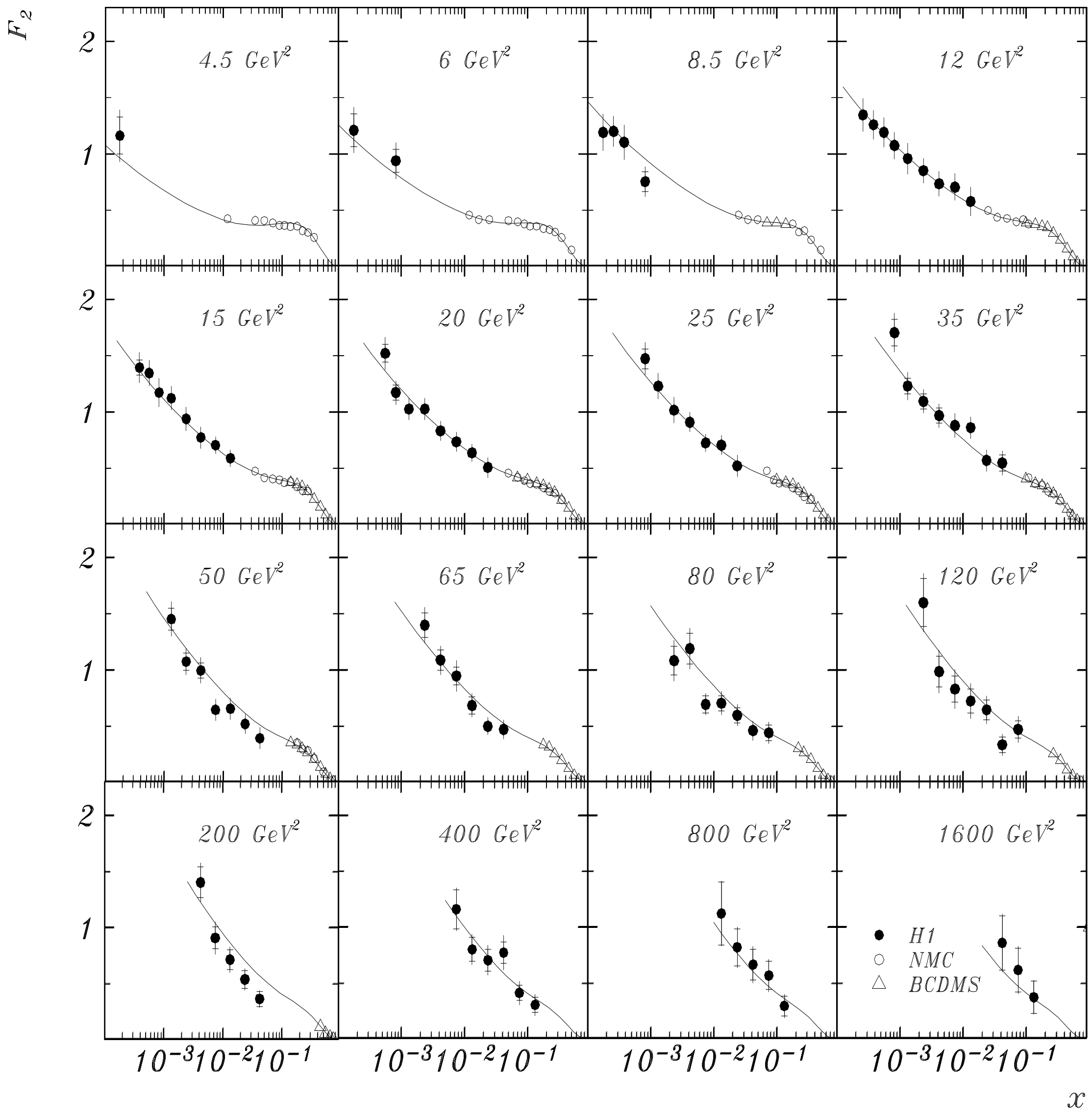

Figure 6: *Measurement of the proton structure function $F_2(x, Q^2)$. The inner error bar is the statistical error. The full error represents the statistical and systematic errors added in quadrature not taking into account the 4.5% systematic error on the luminosity measurement. Open circles and triangles represent NMC and BCDMS measurements, respectively. A smooth transition becomes apparent from the NMC and BCDMS data (open circles and triangles, respectively) to the H1 data. The curves represent a phenomenological fit to all data, see text.*

- An error of up to 3% in the radiative correction due to uncertainties from the hadronic corrections, the cross section extrapolation towards $Q^2 = 0$, second order corrections and the absence of the soft photon exponentiation in the HERACLES Monte Carlo. The accuracy was cross checked by comparing the Monte Carlo estimate [48] with TERAD [49] and also with TERAD supplemented by a leading log higher order calculation [50]. A direct estimate has been made comparing in their overlap region the cross sections derived from the E and the $\Sigma$ methods. Note that most of the Compton events do not contribute here due to the vertex requirement.

- The structure function dependence of the acceptance and bin size corrections which was controlled to better than 3%. The comparison of the different simulation models of the hadronic final state mentioned above was used to assign an additional 3% systematic error to the hadronic methods.

- The uncertainty due to photoproduction background was assumed to be smaller than half the correction applied, i.e. smaller than 5%. This affects only the highest $y$ bins at lower $Q^2$.

- Statistical errors in the Monte Carlo acceptance and efficiency calculations were computed and added quadratically to the systematic error.

- The shifted vertex data sample required to assign a 15% error for the vertex efficiency correction due to lack of experimental statistics. For the higher statistics satellite data the additional luminosity uncertainty implies that both data samples with shifted $z$ vertex position contribute with about the same systematic errors to the final low $Q^2$ structure function measurement.

### 6.3 The Structure Function $F_2(x, Q^2)$

The combination of different methods meant that the largest kinematic range could be covered in an optimum way for the measurement of $F_2(x, Q^2)$. Fig.6 presents the combined structure function data. Data at lower $x$ in the full $Q^2$ range are obtained with the E method. For $y \leq 0.15$ the $\Sigma$ method was used. Double counting of the events was reduced to the level of a few per cent by introducing to the $\Sigma$ analysis in each $Q^2$ bin an $x_e$ cut near to $y = Q^2/sx \sim 0.15$. The results are in good agreement with the previous H1 publication [2].

The structure function values are given in table 4 with their statistical and systematic errors. For low $Q^2$ values, apart from the lower statistics shifted vertex data, the systematic error of $F_2$ is about 10% and twice as large as the statistical error, while for the larger $Q^2$ data the statistical error dominates.

Fig.6 shows that the structure function $F_2$ rises steeply with $x$ decreasing to $x = 1.8 \cdot 10^{-4}$. As can be seen in fig.7 the dependence of $F_2$ on $Q^2$ at fixed $x$ is weak. The violation of scaling appears to be stronger for smaller values of $x$, a trend already observed in fixed target experiments for $x \leq 0.1$. Compared to the recent data from the ZEUS experiment, an extension of the kinematic range is achieved towards low $x$ and low $Q^2$, see fig.7.

The $x$ and $Q^2$ behaviour of $F_2$ can be described by a phenomenological ansatz of the type

$$F_2(x, Q^2) = [a \cdot x^b + c \cdot x^d \cdot (1 + e \cdot \sqrt{x}) \cdot (\ln Q^2 + f \ln^2 Q^2)] \cdot (1 - x)^g. \qquad (7)$$

| $Q^2$ | $x$ | $F_2$ | $\delta_{stat}$ | $\delta_{syst}$ | $R$ |
|---|---|---|---|---|---|
| 4.5 | 0.000178 | 1.16 | 0.17 | 0.16 | 0.34 |
| 6 | 0.000178 | 1.21 | 0.15 | 0.14 | 0.31 |
| 6 | 0.00075 | 0.94 | 0.10 | 0.12 | 0.26 |
| 8.5 | 0.000178 | 1.19 | 0.05 | 0.16 | 0.30 |
| 8.5 | 0.000261 | 1.20 | 0.04 | 0.13 | 0.30 |
| 8.5 | 0.000383 | 1.11 | 0.05 | 0.15 | 0.30 |
| 8.5 | 0.000750 | 0.76 | 0.09 | 0.10 | 0.25 |
| 12 | 0.000261 | 1.35 | 0.06 | 0.13 | 0.29 |
| 12 | 0.000383 | 1.26 | 0.05 | 0.12 | 0.29 |
| 12 | 0.000562 | 1.19 | 0.05 | 0.12 | 0.28 |
| 12 | 0.000825 | 1.08 | 0.04 | 0.11 | 0.28 |
| 12 | 0.00133 | 0.96 | 0.05 | 0.13 | 0.27 |
| 12 | 0.00237 | 0.85 | 0.04 | 0.10 | 0.26 |
| 12 | 0.00421 | 0.74 | 0.04 | 0.10 | 0.25 |
| 12 | 0.00750 | 0.70 | 0.04 | 0.11 | 0.23 |
| 12 | 0.01330 | 0.58 | 0.04 | 0.12 | 0.21 |
| 15 | 0.000383 | 1.40 | 0.07 | 0.12 | 0.28 |
| 15 | 0.000562 | 1.35 | 0.06 | 0.10 | 0.27 |
| 15 | 0.000825 | 1.17 | 0.06 | 0.11 | 0.27 |
| 15 | 0.00133 | 1.13 | 0.05 | 0.10 | 0.26 |
| 15 | 0.00237 | 0.94 | 0.04 | 0.10 | 0.25 |
| 15 | 0.00421 | 0.78 | 0.04 | 0.09 | 0.24 |
| 15 | 0.0075 | 0.71 | 0.04 | 0.07 | 0.22 |
| 15 | 0.0133 | 0.59 | 0.04 | 0.07 | 0.20 |
| 20 | 0.000562 | 1.52 | 0.08 | 0.12 | 0.27 |
| 20 | 0.000825 | 1.17 | 0.07 | 0.08 | 0.26 |
| 20 | 0.00133 | 1.03 | 0.05 | 0.08 | 0.26 |
| 20 | 0.00237 | 1.03 | 0.05 | 0.08 | 0.24 |
| 20 | 0.00421 | 0.83 | 0.04 | 0.08 | 0.23 |
| 20 | 0.0075 | 0.74 | 0.04 | 0.08 | 0.21 |
| 20 | 0.0133 | 0.64 | 0.04 | 0.07 | 0.19 |
| 20 | 0.0237 | 0.51 | 0.05 | 0.08 | 0.16 |
| 25 | 0.000825 | 1.47 | 0.09 | 0.12 | 0.25 |
| 25 | 0.00133 | 1.23 | 0.06 | 0.10 | 0.24 |
| 25 | 0.00237 | 1.02 | 0.06 | 0.10 | 0.23 |
| 25 | 0.00421 | 0.91 | 0.06 | 0.07 | 0.22 |
| 25 | 0.0075 | 0.73 | 0.05 | 0.06 | 0.20 |
| 25 | 0.0133 | 0.71 | 0.05 | 0.07 | 0.18 |
| 25 | 0.0237 | 0.52 | 0.05 | 0.08 | 0.16 |
| 35 | 0.000825 | 1.71 | 0.12 | 0.13 | 0.24 |
| 35 | 0.00133 | 1.23 | 0.07 | 0.11 | 0.23 |
| 35 | 0.00237 | 1.10 | 0.07 | 0.08 | 0.22 |
| 35 | 0.00421 | 0.97 | 0.07 | 0.08 | 0.21 |
| 35 | 0.0075 | 0.88 | 0.06 | 0.09 | 0.19 |
| 35 | 0.0133 | 0.86 | 0.06 | 0.08 | 0.17 |
| 35 | 0.0237 | 0.57 | 0.05 | 0.08 | 0.15 |
| 35 | 0.0421 | 0.55 | 0.07 | 0.08 | 0.12 |
| 50 | 0.00133 | 1.46 | 0.10 | 0.12 | 0.22 |
| 50 | 0.00237 | 1.08 | 0.08 | 0.09 | 0.21 |
| 50 | 0.00421 | 1.00 | 0.07 | 0.09 | 0.20 |
| 50 | 0.0075 | 0.65 | 0.06 | 0.08 | 0.18 |
| 50 | 0.0133 | 0.66 | 0.06 | 0.08 | 0.16 |
| 50 | 0.0237 | 0.52 | 0.05 | 0.07 | 0.14 |
| 50 | 0.0422 | 0.40 | 0.05 | 0.08 | 0.11 |
| 65 | 0.00237 | 1.40 | 0.11 | 0.11 | 0.20 |
| 65 | 0.00421 | 1.09 | 0.09 | 0.09 | 0.19 |
| 65 | 0.00750 | 0.95 | 0.08 | 0.11 | 0.17 |
| 65 | 0.0133 | 0.69 | 0.08 | 0.07 | 0.15 |
| 65 | 0.0237 | 0.50 | 0.06 | 0.05 | 0.13 |
| 65 | 0.0421 | 0.48 | 0.06 | 0.07 | 0.10 |
| 80 | 0.00237 | 1.09 | 0.13 | 0.13 | 0.20 |
| 80 | 0.00421 | 1.19 | 0.14 | 0.12 | 0.18 |
| 80 | 0.00750 | 0.70 | 0.08 | 0.06 | 0.17 |
| 80 | 0.0133 | 0.71 | 0.07 | 0.08 | 0.15 |
| 80 | 0.0237 | 0.60 | 0.07 | 0.07 | 0.12 |
| 80 | 0.0421 | 0.47 | 0.06 | 0.07 | 0.10 |
| 80 | 0.0750 | 0.45 | 0.07 | 0.07 | 0.07 |
| 120 | 0.00237 | 1.60 | 0.21 | 0.15 | 0.18 |
| 120 | 0.00421 | 0.99 | 0.14 | 0.14 | 0.17 |
| 120 | 0.00750 | 0.83 | 0.12 | 0.13 | 0.16 |
| 120 | 0.0133 | 0.73 | 0.11 | 0.11 | 0.14 |
| 120 | 0.0237 | 0.65 | 0.09 | 0.09 | 0.12 |
| 120 | 0.0421 | 0.34 | 0.07 | 0.07 | 0.09 |
| 120 | 0.0750 | 0.48 | 0.08 | 0.09 | 0.07 |
| 200 | 0.00421 | 1.41 | 0.14 | 0.11 | 0.16 |
| 200 | 0.0075 | 0.91 | 0.10 | 0.11 | 0.14 |
| 200 | 0.0133 | 0.72 | 0.09 | 0.08 | 0.13 |
| 200 | 0.0237 | 0.54 | 0.08 | 0.06 | 0.11 |
| 200 | 0.0421 | 0.37 | 0.07 | 0.04 | 0.08 |
| 400 | 0.00750 | 1.16 | 0.17 | 0.08 | 0.13 |
| 400 | 0.0133 | 0.81 | 0.11 | 0.08 | 0.11 |
| 400 | 0.0237 | 0.71 | 0.10 | 0.10 | 0.09 |
| 400 | 0.0421 | 0.78 | 0.10 | 0.12 | 0.07 |
| 400 | 0.0750 | 0.42 | 0.07 | 0.08 | 0.05 |
| 400 | 0.133 | 0.31 | 0.07 | 0.07 | 0.04 |
| 800 | 0.0133 | 1.13 | 0.28 | 0.07 | 0.10 |
| 800 | 0.0237 | 0.82 | 0.16 | 0.08 | 0.08 |
| 800 | 0.0421 | 0.67 | 0.14 | 0.09 | 0.07 |
| 800 | 0.0750 | 0.57 | 0.13 | 0.07 | 0.05 |
| 800 | 0.133 | 0.30 | 0.09 | 0.06 | 0.03 |
| 1600 | 0.0421 | 0.86 | 0.24 | 0.10 | 0.06 |
| 1600 | 0.0750 | 0.62 | 0.19 | 0.09 | 0.04 |
| 1600 | 0.133 | 0.37 | 0.15 | 0.05 | 0.03 |

Table 4: *Proton structure function $F_2(x, Q^2)$ with statistical and systematic errors. All points have an additional scale uncertainty of 4.5% due to the luminosity determination. $Q^2$ is given in GeV$^2$. The values of $R$ result from a calculation based on the QCD prescription using the MRSH parton distribution parametrization.*

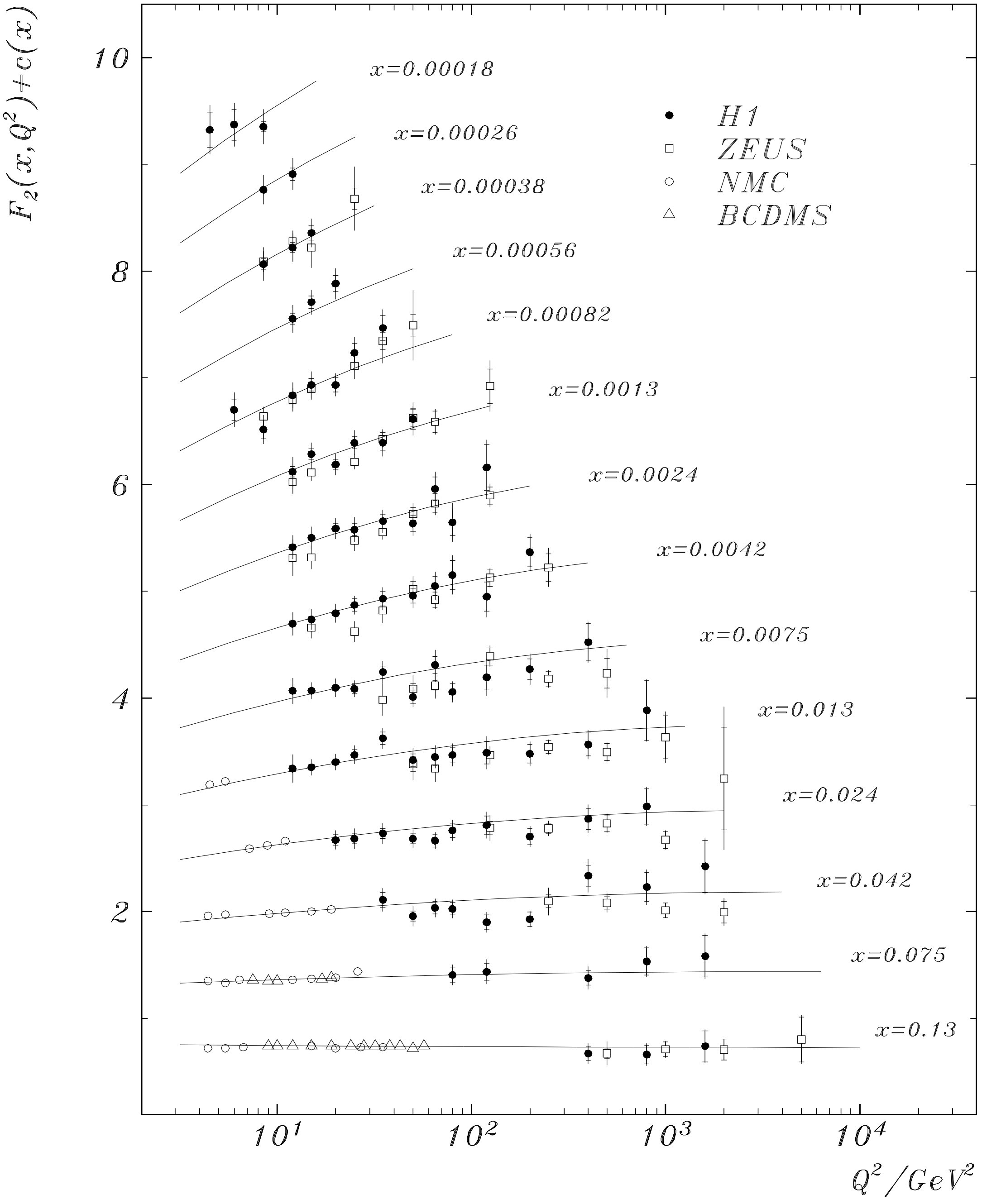

Figure 7: *Measurement of the proton structure function* $F_2(x, Q^2)$*. The H1 data (closed points) is consistent with the ZEUS result (open squares) but extends further towards low* $x$ *and also to lower* $Q^2$ *in different* $x$ *bins. With a small correction the ZEUS* $F_2$ *data was shifted to the H1* $x$ *values by using the parametrization given in [8]. The curves represent a phenomenological fit to the H1, NMC and BCDMS data, see text. The* $F_2$ *values are plotted with all but normalization errors in a linear scale adding a term* $c(x) = 0.6(i_x - 0.4)$ *to* $F_2$ *where* $i_x$ *is the bin number starting at* $i_x = 1$ *for* $x = 0.13$.

The dependence of that parametrization reflects the following observations: i) it is constructed to describe the $F_2$ data from BCDMS, NMC and H1 which together cover a range of 4 orders of magnitude in $x$ and $Q^2$; ii) for large $x$, $F_2$ is known to vanish like $(1-x)^g$ with the parameter $g$ close to 3 as determined by quark counting rules; iii) due to momentum conservation the integral over $F_2$ is nearly independent of $Q^2$. Thus the rise of $F_2$ with $Q^2$ at low $x$ must be compensated by a decrease of $F_2$ with $Q^2$ at large $x$. As a consequence, at $x = x_o$, $F_2$ is independent of $Q^2$. This demands that a polynomial term be included in front of the $Q^2$ dependent part which has been chosen to be $(1 + e\sqrt{x})$. Since $x_o$ in the fixed target experiments has been determined to be around 0.12, the parameter $e$ should be close to $-3$; iv) the $Q^2$ dependence of $F_2$ is expected to be logarithmic. The attempt to describe the $Q^2$ evolution over almost four orders of magnitude, from $Q^2 \geq 4$ GeV$^2$ to $Q^2 \leq 2000$ GeV$^2$, requires a quadratic term in $\ln Q^2$. Fits were performed with and without that term present; v) finally the introduction of a term in eq.7 is required which is independent of $Q^2$.

| a | b | c | d | e | f | g |
|---|---|---|---|---|---|---|
| 3.07 | 0.75 | 0.14 | -0.19 | -2.93 | -0.05 | 3.65 |

Table 5: *Parameters of a phenomenological fit to the proton structure function data from this experiment combined with $F_2^p$ from the NMC and the BCDMS experiments. The parametrization is valid for* $4$ *GeV*$^2 < Q^2 < 2000$ *GeV*$^2$, $10^{-4} < x < 1$ *and* $Q^2 < x \cdot 10^5$ *GeV*$^2$.

The result of the fit to the H1, NMC and BCDMS data is shown as functions of $x$ and $Q^2$ in figs. 6 and 7. The parameter values are quoted in table 5. In the fit the statistical and systematic errors of all quoted $F_2$ values were added in quadrature and the relative normalizations were not allowed to vary. The fit provides a valid description of all data from the experiments considered here with a $\chi^2/dof$ of 1.34. For the H1 data alone the parametrization gives a $\chi^2/dof$ of 0.85. The parameters $g$ and $e$ come out as expected to be close to $\pm 3$. From the result $e = -2.93$ one finds that the slope of $F_2$ with $\ln Q^2$ changes sign at $x_o \simeq 0.12$. The obtained parameter $f$ of eq.7 is small, i.e. the $\ln^2 Q^2$ term amounts to about a 20% correction to the linear behaviour at $Q^2$ values between 20 and 50 GeV$^2$. If the $\ln^2 Q^2$ term in eq.7 is neglected the $\chi^2/dof$ increases from 1.34 to 1.55. An interesting result of the parametrization is the interplay between the first and the second term with an exponential $x$ dependence. The second term dominates at low $x$. The value of the power $d$ is found to be correlated with the presence of the $\ln^2 Q^2$ term. If the latter is neglected the fit suggests an $x$ dependence at low $x$ which is somewhat steeper, namely $d = -0.23$ instead of $d = -0.19$.

In fig.8 the $F_2$ behaviour is illustrated for different $Q^2$ values as a function of $W$ which is the invariant mass of the virtual photon-proton ($\gamma^* p$) system,

$$W = \sqrt{Q^2 \cdot (1/x - 1) + M_p^2} \simeq \sqrt{Q^2/x} \quad (8)$$

at low $x$. Here $M_p$ is the proton mass. Since $Q^2/x = sy$, the HERA experiments reach much larger $W$ values than the previous DIS or real photoproduction experiments. $F_2$ is displayed for the high statistics nominal vertex data and only for those $Q^2$ bins where $x$ below $10^{-3}$ was reached, i.e. from 8.5 GeV$^2$ to 35 GeV$^2$. The rise of $F_2$ at low $x$ corresponds directly to a rise of $F_2$ with $W$. This can be interpreted as a strong rise of the $\gamma^* p$ total cross section, since for $x << 1$ and in the region where effects due to $Z^o$ boson exchange can be neglected,

$$\sigma_{tot}(\gamma^* p) \simeq \frac{4\,\pi^2 \alpha}{Q^2} F_2(W, Q^2). \quad (9)$$

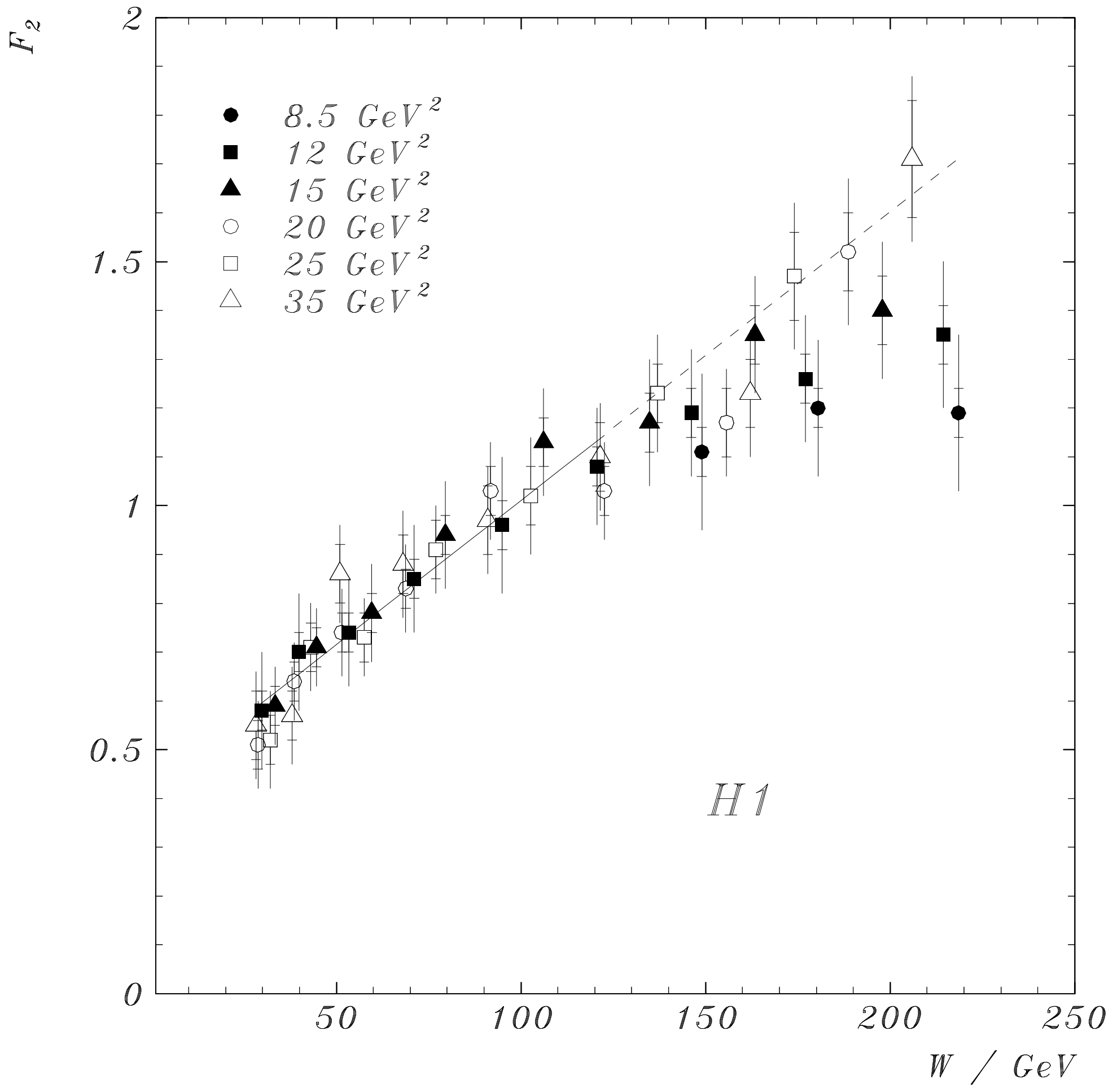

Figure 8: *Low $Q^2$ structure function data of H1 plotted as functions of the invariant mass $W$ of the $\gamma^* p$ system. The straight line is a fit $F_2(W, Q^2) = 0.0058\ W/GeV + 0.42$ for $x > 0.001$ which extends up to $W \simeq 130$ GeV where $F_2$ scales for the $Q^2$ range considered. Data for $Q^2 \geq 80$ GeV$^2$ have a similar slope versus $W$ but are below the lower $Q^2$ data.*

Three observations are made: i) the rise of $F_2$ with $W$ is stronger than the one observed for $W$ between 20 and 200 GeV of the total photoproduction cross section [23, 51], in which the photon is real. This behaviour for the total cross section of off-shell particle scattering was expected [52] and discussed for previous DIS experiments in [53]; ii) for $W$ values below $\simeq 130$ GeV, corresponding to $x > 10^{-3}$, all the measured points cluster in a narrow band which is well reproduced by a straight line fit, i.e. $F_2(W, Q^2)$ scales in the $Q^2$ range considered. This observation is consistent with the recent ZEUS measurement [8]; iii) the extrapolation of the straight line fit (dashed line in fig.8) into the higher $W$ region, $W \geq 130$ GeV, reproduces the data at $Q^2 \geq 20$ GeV$^2$ but the $F_2$ data at lower $Q^2$ and $x \leq 10^{-3}$ appear to deviate systematically from this linear behaviour. This trend is confirmed, though with less precision, by the lowest $Q^2$ measurements obtained with the shifted vertex data.

The behaviour of $F_2$ at very low $x$ requires still more precision data and extended coverage than were available to this analysis. A more theoretical discussion of the $F_2$ data presented in this paper is deferred to a forthcoming analysis in the framework of perturbative Quantum Chromodynamics.

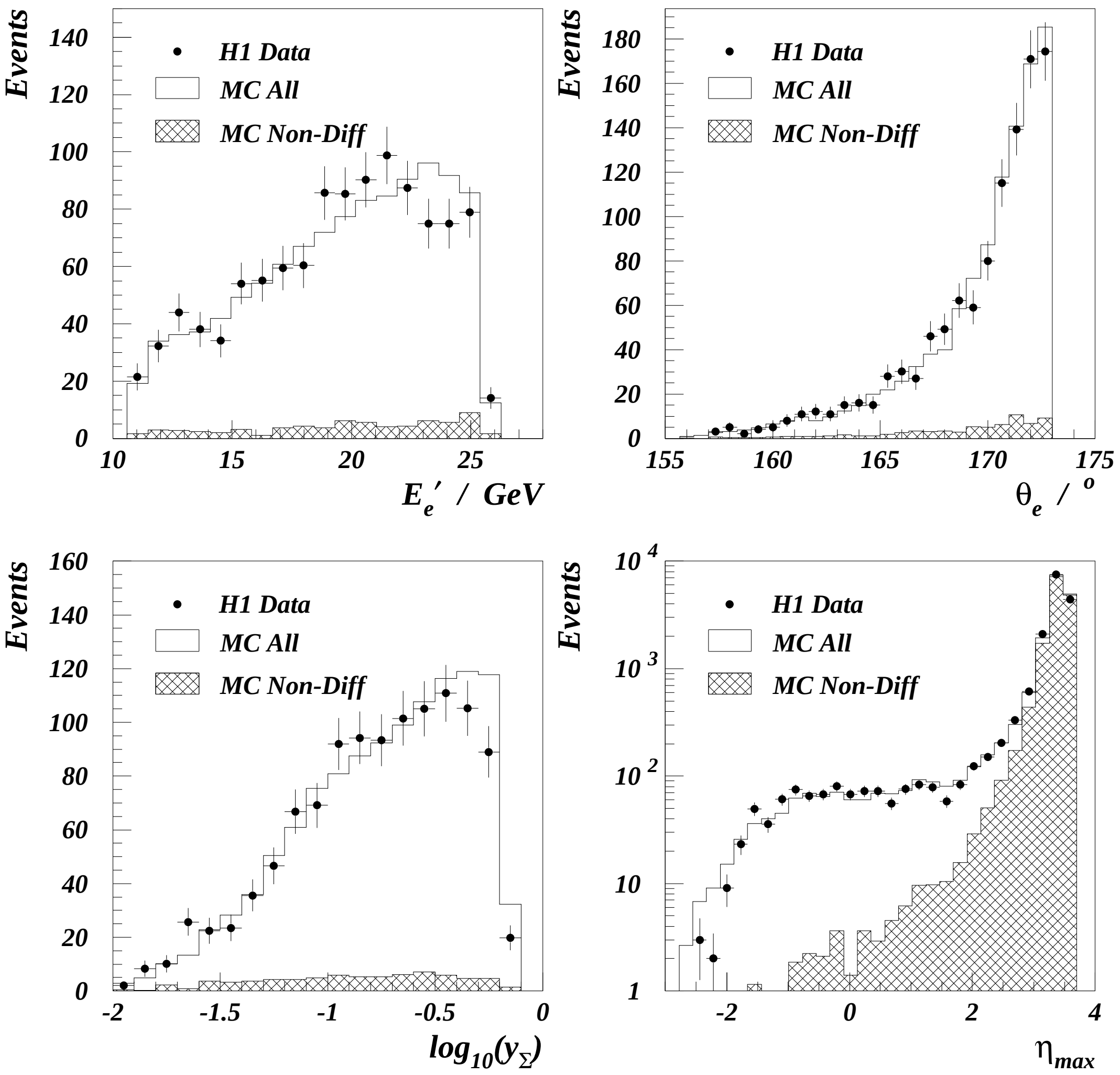

Figure 9: *$E_e'$, $\theta_e$ and $y_\Sigma$ distributions for DIS events with $\eta_{max} < 1.8$ and the $\eta_{max}$ distribution for all DIS selected events. The Monte Carlo simulations are described in the text.*

# 7 Diffractive Contribution to $F_2(x, Q^2)$

In the previous H1 measurement of $F_2(x, Q^2)$ [2], it has been observed that for about 6% of the deep inelastic events there was no significant energy deposition in the forward region. The angular acceptance of the LAr calorimeter is limited to polar angles $\theta > 3^o$ corresponding to a maximum measurable pseudo-rapidity, $\eta = -\ln\left(\tan\frac{\theta}{2}\right)$, of 3.6. The $\eta$ value of the most forward cluster [39] with energy greater than 400 MeV is defined to be $\eta_{max}$. The events with $\eta_{max} < 1.8$ amount to 6.3% of the low $Q^2$ data. A first H1 analysis of the rapidity gap events collected in 1993 has been reported in [10]. There it was shown that these events may be described by diffractive scattering processes where the virtual photon probes a colourless component of the proton, the Pomeron. The kinematics of deep inelastic diffractive scattering leads to a relationship [54] between the minimum size of the rapidity gap and $x_{I\!P/p}$, the momentum fraction of the proton carried by the Pomeron, such that

$$\Delta\eta = \eta_{proton} - \eta_{max} \geq ln\,\frac{1}{x_{I\!P/p}} \qquad \text{with} \qquad x_{I\!P/p} = x \cdot \left(1 + \frac{M_X^2}{Q^2}\right) \tag{10}$$

where $M_X$ is the invariant mass of the $\gamma^* I\!P$ system. This relation implies that a selection requiring $\eta_{max} < 1.8$ introduces a limit of $x_{I\!P/p} \leq 0.01$ on the data [55]. The diffractive event sample contains both interactions with a scattered proton, and with a proton dissociative system. The analysis uses the same data sample and selection criteria as for the $F_2(x,Q^2)$ measurement with the additional event selection of $\eta_{max} < 1.8$.

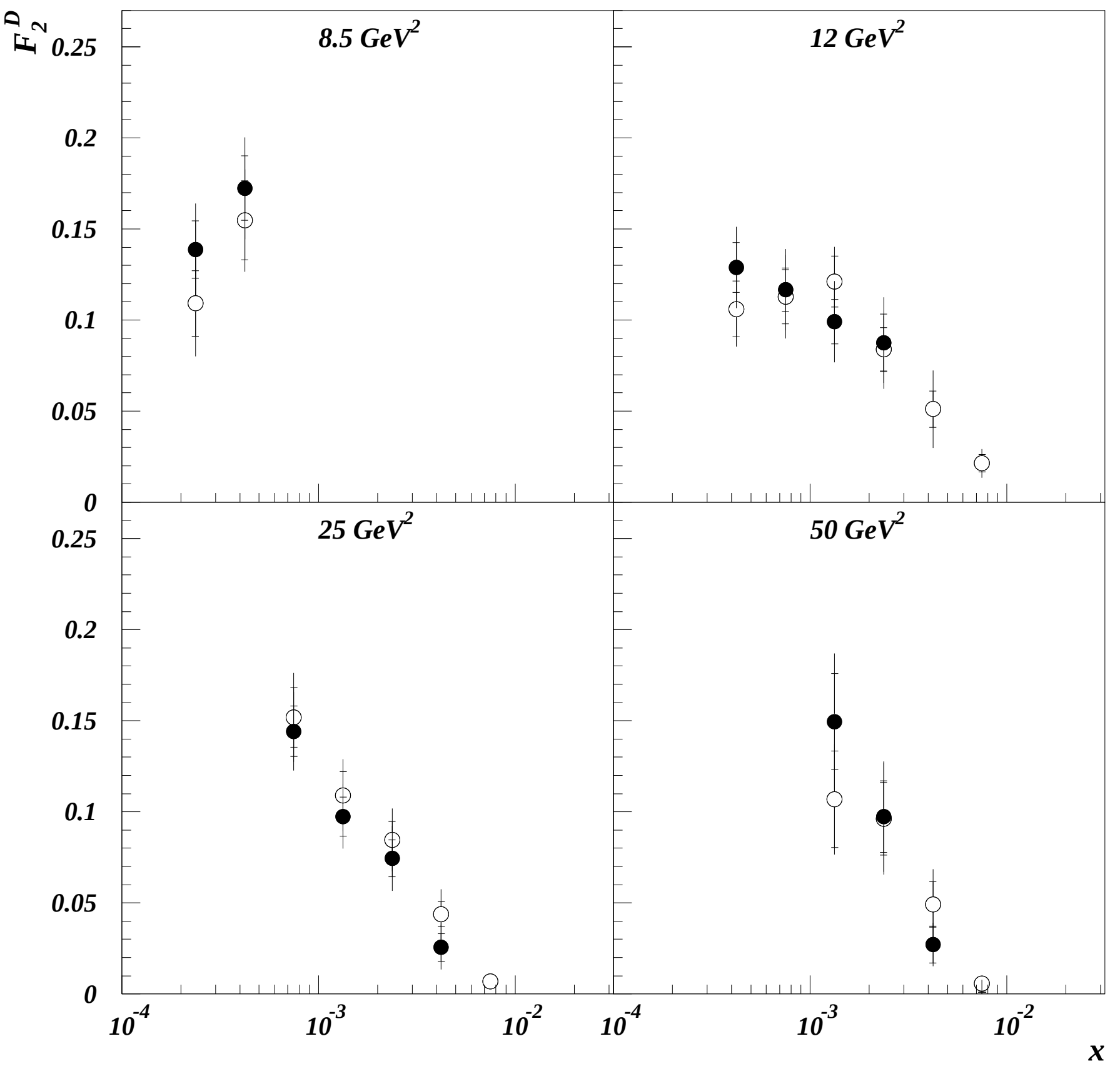

Figure 10: *Measurement of the diffractive contribution* $F_2^D(x,Q^2)$ *to* $F_2(x,Q^2)$ *for* $x_{I\!P/p} < 10^{-2}$ *and different* $Q^2$ *values with the E method (closed circles) and the* $\Sigma$ *method (open circles). The inner error bar is the statistical error. The full error bar represents the statistical and systematic errors added in quadrature, not including the systematic error of the luminosity measurement of 4.5%.*

Following [56] the diffractive contribution $F_2^D(x,Q^2)$ to $F_2(x,Q^2)$ for $x_{I\!P/p} < 0.01$ can be

defined as

$$F_2^D(x,Q^2) = \frac{1}{\kappa} \cdot \int_x^{10^{-2}} \int_{t_{min}}^{\infty} \frac{d^4\sigma^D}{dx\, dQ^2\, dx_{I\!P/p}\, dt} dt\, dx_{I\!P/p} \tag{11}$$

with the diffractive cross section $\sigma^D$ and the kinematic factor $\kappa$ defined as in eq.6, setting $R = 0$. Here $t$ is the momentum transfer squared between the incident and scattered proton or proton dissociative system. To correct for the acceptance and for effects due to finite resolution, the RAPGAP Monte Carlo program [46] was used assuming a hard structure function for the Pomeron $\propto x_{i/I\!P}(1-x_{i/I\!P})$ with the same amount of quark and gluon induced events. Here $x_{i/I\!P}$ is the fraction of the Pomeron momentum carried by the parton $i$ which is assumed to interact with the virtual photon. A contribution from the elastic production of light vector mesons ($\rho(770), \omega(783), \phi(1020)$), consistent with that observed in [10], was added to the RAPGAP Monte Carlo events [57] amounting to 10% of the selected rapidity gap sample. This simulation describes the data well as shown in fig.9. Here the $E'_e, \theta_e, y_\Sigma$ distributions for DIS events with $\eta_{max} < 1.8$ and the $\eta_{max}$ distribution for all DIS events are shown for both data and Monte Carlo simulation.

The $F_2^D$ contributions derived from the electron method and from the $\Sigma$ method are shown in fig.10. Both measurements agree well within the errors in the regions of overlap. As for the measurement of $F_2$ the $\Sigma$ method was taken for $y < 0.15$ and the electron method for larger $y$. The resulting values for $F_2^D$ are given in table 6. The systematic errors were calculated as for the inclusive $F_2(x,Q^2)$ measurement taking into account the uncertainties specific to the rapidity gap event analysis [55].

| $Q^2/\mathrm{GeV}^2$ | $x$ | $F_2^D$ | $\delta_{stat}$ | $\delta_{syst}$ |
|---|---|---|---|---|
| 8.5 | 0.000237 | 0.134 | 0.016 | 0.020 |
| 8.5 | 0.000421 | 0.172 | 0.018 | 0.022 |
| 12 | 0.000421 | 0.129 | 0.014 | 0.017 |
| 12 | 0.000750 | 0.117 | 0.012 | 0.019 |
| 12 | 0.00133 | 0.121 | 0.014 | 0.013 |
| 12 | 0.00237 | 0.084 | 0.012 | 0.014 |
| 12 | 0.00421 | 0.051 | 0.010 | 0.019 |
| 12 | 0.00750 | 0.021 | 0.0047 | 0.0062 |
| 25 | 0.000750 | 0.144 | 0.014 | 0.017 |
| 25 | 0.00133 | 0.097 | 0.011 | 0.014 |
| 25 | 0.00237 | 0.085 | 0.010 | 0.014 |
| 25 | 0.00421 | 0.044 | 0.007 | 0.012 |
| 25 | 0.00750 | 0.0069 | 0.0020 | 0.0013 |
| 50 | 0.00133 | 0.150 | 0.026 | 0.026 |
| 50 | 0.00237 | 0.097 | 0.020 | 0.023 |
| 50 | 0.00421 | 0.050 | 0.012 | 0.015 |
| 50 | 0.00750 | 0.0056 | 0.0032 | 0.0019 |

Table 6: *Diffractive contribution $F_2^D(x,Q^2)$ to $F_2(x,Q^2)$ for $x_{I\!P/p} < 0.01$. All points have an additional scale error of 4.5% due to the uncertainty of the luminosity measurement.*

By construction $F_2^D$ decreases to zero at $x = 10^{-2}$. Comparison of the values of $F_2^D(x,Q^2)$ with $F_2(x,Q^2)$ for $x < 10^{-3}$ shows that the former contributes about 10% to the proton structure function. Therefore diffraction with $x_{I\!P/p} \leq 10^{-2}$ cannot account for the steep increase of $F_2$ with decreasing $x$. The $Q^2$ dependence of $F_2^D(x,Q^2)$ at fixed $x$ is flat and not significantly different from the $Q^2$ dependence of the $F_2(x,Q^2)$. An analysis of diffractive events in terms of structure functions will be given in more detail in a forthcoming publication of H1 [58].

# 8 Summary

A measurement has been presented of the proton structure function $F_2(x, Q^2)$ in deep inelastic electron-proton scattering at HERA with data taken in the running period of 1993. The integrated luminosity is 0.271 pb$^{-1}$ which represents a tenfold increase in statistics compared to the first $F_2$ publication of H1. The structure function measurement includes data from different detector components and running configurations. Low $Q^2$ values are reached using data with the $ep$ interaction vertex shifted in $z$ from the nominal position. The data cover a kinematic range for $Q^2$ between 4.5 and 1600 GeV$^2$ and $x$ between $1.8 \cdot 10^{-4}$ and 0.13.

The $F_2$ values presented are obtained using different methods to reconstruct the inclusive scattering kinematics. At high values of the scaling variable $y \geq 0.15$, due to its superior resolution, the electron method is used which is based on the scattered electron energy and angle. Lower $y$ values are covered with the $\Sigma$ method which combines electron and hadronic information to reduce radiative corrections and calibration errors. For comparison and cross-checks, $F_2$ data obtained with the double angle method are also presented.

The measured structure function has statistical and systematic errors considerably smaller than the previous result. For the first time, the HERA structure function measurement extends to the kinematical region of the high precision, fixed target experiments revealing a smooth transition between the BCDMS, NMC and H1 results.

A distinct rise is observed of the structure function with decreasing $x$ at fixed $Q^2$. Around $x \sim 10^{-3}$ the decrease of $x$ by an order of magnitude amounts to a rise of $F_2$ of about a factor two. This rise cannot be explained by the rapidity gap events because they contribute only about 10% of $F_2$. They are used to measure for the first time the diffractive contribution $F_2^D$ to the proton structure function for $x_{I\!P/p}$ smaller than $10^{-2}$. This contribution exhibits no significant $Q^2$ dependence.

The observed $Q^2$ behaviour is consistent with the expected scaling violations, i.e. a weak rise of $F_2$ with increasing $Q^2$ for $x < 0.1$. A parametrization is given of the proton structure function data from this experiment combined with the data from the NMC and BCDMS experiment describing $F_2(x, Q^2)$ over almost four orders of magnitude in $x$ and $Q^2$. If $F_2$ at low $Q^2$ is analyzed as a function of the virtual photon-proton mass $W$, deviations from a linear behaviour become visible for $W \geq 130$ GeV which deserve a more precise analysis with forthcoming data.

**Acknowledgements**

We are very grateful to the HERA machine group whose outstanding efforts made this experiment possible. We acknowledge the support of the DESY technical staff. We appreciate the big effort of the engineers and technicians who constructed and maintained the detector. We thank the funding agencies for financial support of this experiment. We wish to thank the DESY directorate for the support and hospitality extended to the non-DESY members of the collaboration.

# References

[1] for a recent review see: J. Feltesse, DAPNIA-SPP-94-35(1994), Invited talk at the 27. International Conference on High Energy Physics, Glasgow, Scotland, 1994.

[2] H1 Collab., I. Abt et al., Nucl. Phys. **B407** (1993) 515.

[3] ZEUS Collab., M. Derrick et al., Phys. Lett. **B316** (1993) 412.

[4] A. De Rújula et al, Phys. Rev. **D10** (1974) 1649.

[5] Yu. L. Dokshitzer, Sov. Phys. JETP **46** (1977) 641;
V. N. Gribov and L.N. Lipatov, Sov. J. Nucl. Phys. **15** (1972) 438 and 675;
G. Altarelli and G. Parisi, Nucl. Phys. **B126** (1977) 297.

[6] E. A. Kuraev, L. N. Lipatov and V. S. Fadin, Sov. Phys. JETP **45** (1977) 199;
Y. Y. Balitsky and L.N. Lipatov, Sov. J. Nucl. Phys. **28** (1978) 822.

[7] L. V. Gribov, E. M. Levin and M. G. Ryskin, Phys. Rep. **100** (1983) 1;
A. H. Mueller and N. Quiu, Nucl. Phys. **B268** (1986) 427.

[8] ZEUS Collab., M. Derrick et al., DESY 94-143 (1994), submitted to Z. Phys. **C**.

[9] ZEUS Collab., M. Derrick et al., Phys. Lett. **B315** (1993) 481,
Phys. Lett. **B332** (1994) 228 and Phys. Lett. **B338** (1994) 483.

[10] H1 Collab., T. Ahmed et al., Nucl. Phys. **B429** (1994) 477.

[11] A. Blondel and F. Jacquet, Proceedings of the study of an *ep* facility for Europe, ed. U.Amaldi, DESY 79/48 (1979) p.391.

[12] J. Blümlein and M. Klein, Proceedings of the Snowmass Workshop "The Physics of the Next Decade", ed. R. Craven, (1990) p.549.

[13] S. Bentvelsen et al., Proceedings of the Workshop Physics at HERA, Vol. 1, eds. W. Buchmüller and G. Ingelman, DESY (1992) p.23;
C. Hoeger, ibid. p.43.

[14] U. Bassler and G. Bernardi, DESY 94-231(1994), subm. to Nucl. Instr. and Meth.

[15] H1 Collab., I. Abt et al., DESY 93-103 (1993).

[16] H1 Calorimeter Group, B. Andrieu et al., Nucl. Instr. and Meth. **A336** (1993) 460.

[17] H1 Collab., T. Ahmed et al., to be published.

[18] H1 Collab., I. Abt et al., Phys. Lett. **B340** (1994) 205.

[19] C. Leverenz, Ph.D. Thesis , University of Hamburg (1995), in litt.

[20] E. Peppel, Ph.D. Thesis , University of Hamburg (1994).

[21] N. H. Brook, A. De Roeck and A. T. Doyle, RAYPHOTON 2.0, Proceedings of the Workshop Physics at HERA, vol. 3, eds. W. Buchmüller, G. Ingelman, DESY (1992) p.1453.

[22] H. U. Bengtsson and T. Sjöstrand, Computer Phys. Comm. **46** (1987) 43.

[23] H1 Collab., T. Ahmed et al., Phys. Lett. **B299** (1993) 374;
M. Erdmann et al., DESY 93-077 (1993);
S. Levonian, Proc. of the 28th Rencontre De Moriond, Les Arcs, France, 1993, p.529.

[24] H1 Collab., I. Abt et al., Phys. Lett. **B328** (1994) 176.

[25] S. Reinshagen, Ph.D. Thesis , University of Hamburg (1995), in litt.

[26] D. Neyret, Ph.D. Thesis , University VI of Paris (1995), in litt.

[27] K. Müller, Ph.D. Thesis , University of Zurich (1994).

[28] G. A. Schuler and H. Spiesberger, Proceedings of the Workshop Physics at HERA, vol. 3, eds. W. Buchmüller, G. Ingelman, DESY (1992) p.1419-1432.

[29] A. Kwiatkowski, H. Spiesberger and H.-J. Möhring, Computer Phys. Comm. **69** (1992) 155.

[30] G. Ingelman, Proceedings of the Workshop Physics at HERA, vol. 3, eds. W. Buchmüller, G. Ingelman, DESY (1992) p.1366.

[31] A. D. Martin, W. J. Stirling and R. G. Roberts, Proceedings of the Workshop on Quantum Field Theory Theoretical Aspects of High Energy Physics, eds. B. Geyer and E. M. Ilgenfritz (1993) 11-26.

[32] L. Lönnblad, Computer Phys. Comm. **71** (1992) 15.

[33] H1 Collab., I. Abt et al., Z. Phys. **C63** (1994) 377.

[34] ZEUS Collab., M. Derrick et al., Z. Phys. **C59** (1993), 231.

[35] G. Marchesini et al., Computer Phys. Comm. **67** (1992) 465.

[36] R. Brun et al., GEANT3 User's Guide, CERN–DD/EE 84–1, Geneva (1987).

[37] C. Brune, Ph.D. Thesis , University of Heidelberg (1995), in litt.

[38] H1 Calorimeter Group, B. Andrieu et al., Nucl. Instr. and Meth. **A350** (1994) 57.

[39] H1 Calorimeter Group, B. Andrieu et al., Nucl. Instr. and Meth. **A336** (1993) 499.

[40] H. P. Wellisch et al., MPI-PhE/94-03(1994).

[41] U. Bassler, Ph.D. Thesis , University VI of Paris (1993).

[42] G. Bernardi and W. Hildesheim, Proceedings of the Workshop Physics at HERA, vol. 1, eds. W. Buchmüller, G. Ingelman, DESY (1992) p. 79.

[43] NMC Collab., P. Amaudruz et al., Phys. Lett. **B259** (1992) 159.

[44] BCDMS Collab., A. C. Benvenuti et al., Phys.Lett. **B237** (1990) 592.

[45] G. Altarelli and G. Martinelli, Phys. Lett. **76B** (1978) 89.

[46] H. Jung, DESY 93-182 (1993), submitted to Comput. Phys. Commun.

[47] U. Stößlein, Ph.D. Thesis , Humboldt University of Berlin (1995), in litt.

[48] U. Obrock, Ph.D. Thesis , University of Dortmund (1994).

[49] A. Akhundov et al., DESY 94-115 (1994).

[50] J. Blümlein, Proceedings of the Workshop Physics at HERA, vol. 3, eds. W. Buchmüller, G. Ingelman, DESY (1992) p.1270 and DESY 94-044, Z. Phys. **C** (1994) in press.

[51] ZEUS Collab., M. Derrick et al., Phys. Lett. **B293** (1992) 465, DESY 94-032 (1994), to be published in Z. Phys. **C**.

[52] C. López and F. J. Ynduráin, Phys. Rev. Lett. **44** (1980) 1118.

[53] A. Levy and U. Maor, Phys. Lett. **B182** (1986) 108.

[54] E. L. Berger et al., Nucl. Phys. **B286** (1987) 704.

[55] J. P. Phillips, Ph.D. Thesis, University of Manchester (1995), in litt.

[56] G. Ingelman and K. Prytz, Z. Phys. **C58** (1993) 285.

[57] B. List, Diploma Thesis, Techn. Univ. Berlin, 1993 (unpublished).

[58] H1 Collab., T. Ahmed et al., to be published.

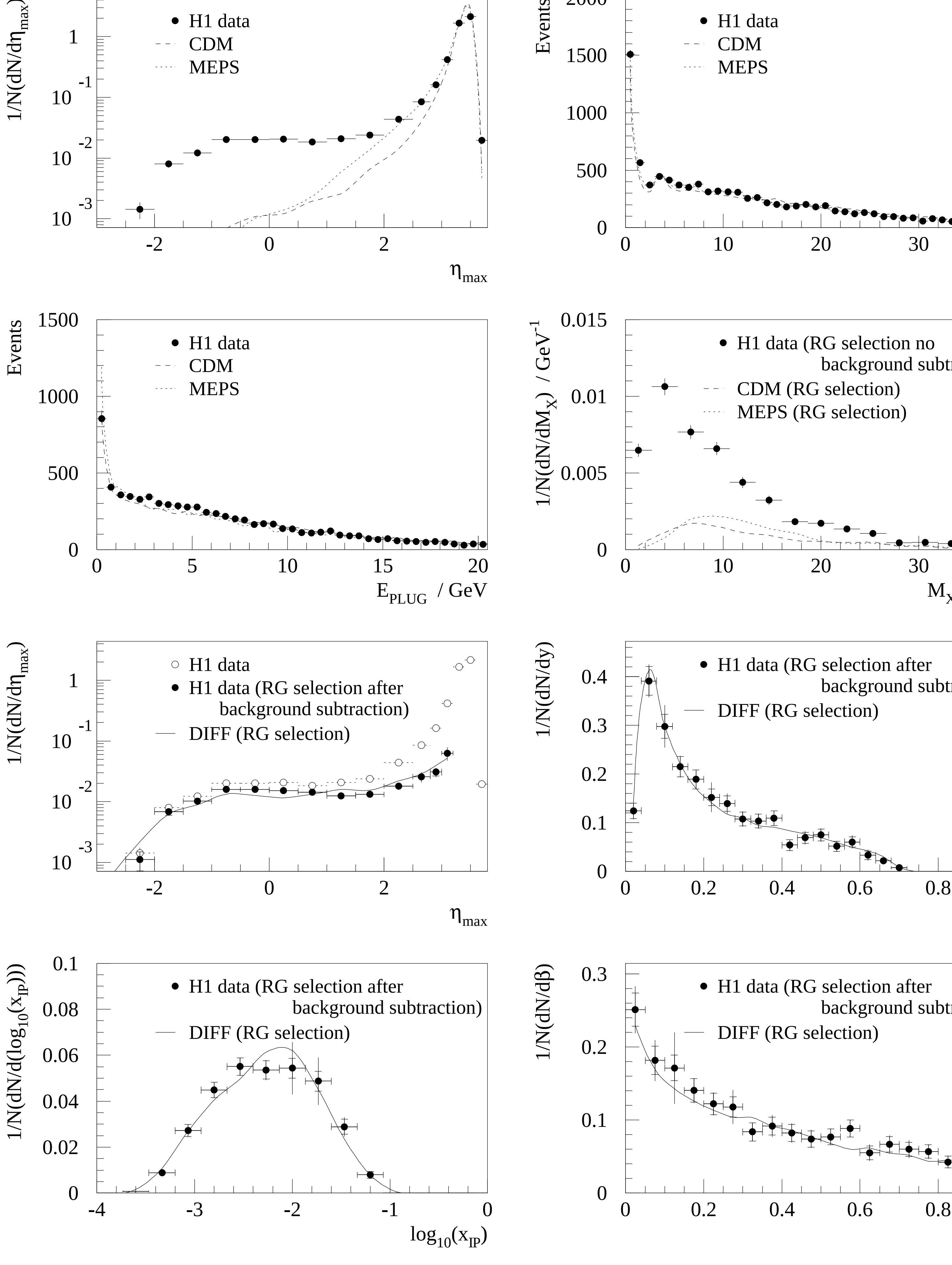

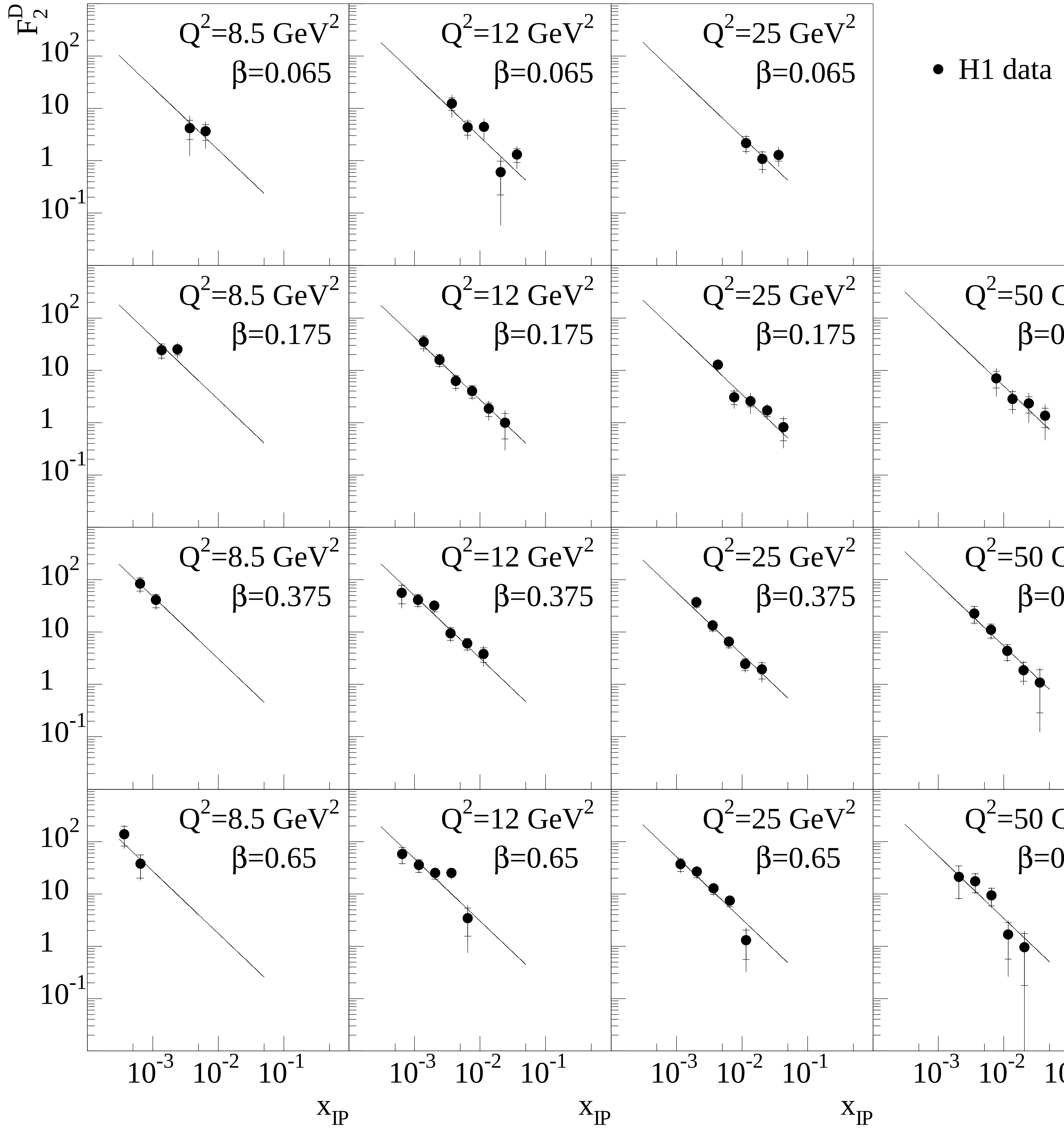

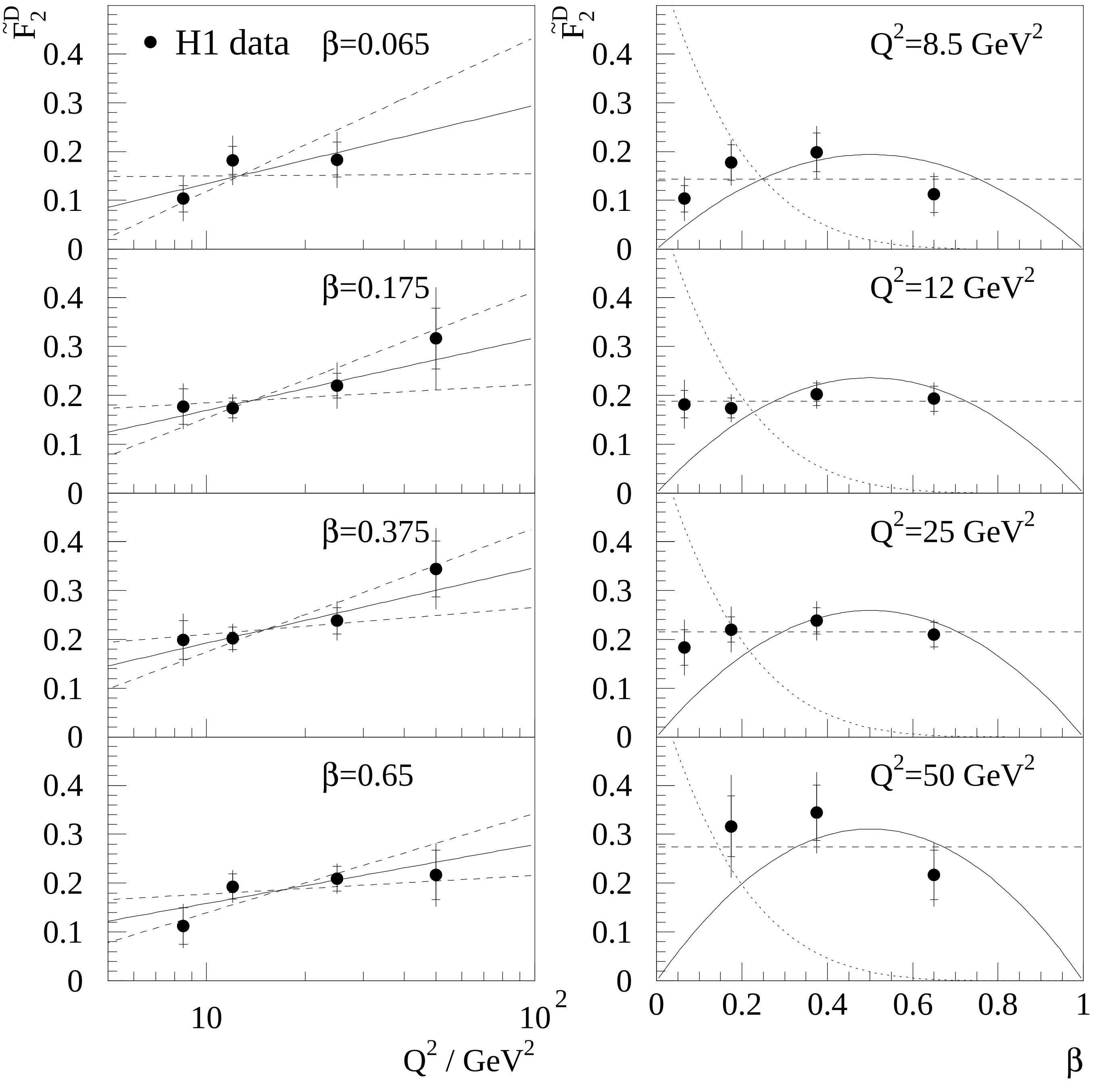